\documentclass[a4paper,aps,prd,preprintnumbers]{revtex4-1}
\pdfoutput=1
\usepackage[utf8]{inputenc}
\usepackage{amsmath}
\usepackage{slashed}
\usepackage{graphicx} 
\usepackage{graphics,subfigure} 
\usepackage{hyperref} 
\usepackage{xcolor}
\usepackage{ulem}
\usepackage{afterpage}

\newcommand{\met}{\ensuremath{\slashed{E}_T}}

\newbox\charbox 
\newbox\slabox 
\def\s#1{{      
		\setbox\charbox=\hbox{$#1$} 
		\setbox\slabox=\hbox{$/$} 
		\dimen\charbox=\ht\slabox 
		\advance\dimen\charbox by -\dp\slabox 
		\advance\dimen\charbox by -\ht\charbox 
		\advance\dimen\charbox by \dp\charbox 
		\divide\dimen\charbox by 2 
		\raise-\dimen\charbox\hbox to \wd\charbox{\hss/\hss} 
		\llap{$#1$} 
}}

\def\lsim{\raise0.3ex\hbox{$\;<$\kern-0.75em\raise-1.1ex\hbox{$\sim\;$}}}
\def\gsim{\raise0.3ex\hbox{$\;>$\kern-0.75em\raise-1.1ex\hbox{$\sim\;$}}}
\begin{document}
	\title{Confronting the neutralino and chargino sector of the NMSSM to the multi-lepton searches at the LHC}
	
	\preprint{IFT-UAM/CSIC-18-119, BONN-TH-2018-15 }
	\author{Florian Domingo} 
	\email[]{florian.domingo@csic.es} 
	\affiliation{Bethe Center for Theoretical Physics \& Physikalisches Institut der Universit\"at Bonn, Nu\ss allee 12, 53115 Bonn, Germany}
	\affiliation{Instituto de F\'isica Te\'orica (UAM/CSIC), Universidad Aut\'onoma de Madrid, Cantoblanco, 28049 Madrid, Spain}
	
	\author{Jong Soo Kim} 
	\email[]{jongsoo.kim@tu-dortmund.de} 
	\affiliation{National Institute for Theoretical Physics and 
		School of Physics, University of the Witwatersrand, Johannesburg, Wits 2050, South Africa}  
	
	\author{V\'ictor Mart\'in Lozano} 
	\email[]{lozano@physik.uni-bonn.de} 
	\affiliation{Bethe Center for Theoretical Physics \& Physikalisches Institut der Universit\"at Bonn, Nu{\ss}allee 12, 
		53115 Bonn, Germany, 
		53115 Bonn, Germany} 
	
	\author{Pablo Mart\'in-Ramiro} 
	\email[]{pmartin.ramiro@predoc.uam.es} 
	\affiliation{Instituto de F\'isica Te\'orica UAM-CSIC, Universidad Aut\'onoma de Madrid,
		Cantoblanco, 28049 Madrid, Spain}

	\author{Roberto Ruiz de Austri} 
	\email[]{rruiz@ific.uv.es} 
	\affiliation{Instituto de F\'isica Corpuscular CSIC--UV, C/ Catedr\'atico Jos\'e Beltr\'an 2, 46980 Paterna, Valencia, Spain}

	\begin{abstract}
		We test the impact of the ATLAS and CMS multi-lepton searches performed at the LHC with 8 as well as 13~TeV center-of-mass energy (using only the pre-2018 results) on the chargino and neutralino sector of the NMSSM. Our purpose consists in analyzing the actual reach of these searches for a full model and in emphasizing effects beyond the MSSM that affect the performance of current (MSSM-inspired) electroweakino searches. To this end, we consider several scenarios characterizing specific features of the NMSSM electroweakino sector. We then perform a detailed collider study, generating Monte-Carlo events through \texttt{Pythia} and testing against current LHC constraints implemented in the public tool \texttt{CheckMATE}. We find e.g.\ that SUSY decay chains involving intermediate singlino or Higgs-singlet states can modify the naive MSSM-like picture of the constraints by inducing final-states with softer or less-easily identifiable SM particles -- reversely, a compressed configuration with singlino NLSP occasionally induces final states that are rich with photons, which could provide complementary search channels. 
	\end{abstract}
	
	\maketitle
	
	\section{Introduction}
	After several years of operation, the LHC has been placing limits on the production of particles beyond the Standard Model (SM), thus constraining several scenarios of new physics. The most severe limits apply on the production of colored particles	and typically exclude most candidates with mass well above the TeV range \cite{Aaboud:2017vwy,Sirunyan:2018vjp}, with notable caveat and exceptions however \cite{Drees:2012dd,Drees:2015aeo,Fan:2011yu,Buckley:2014fqa,Aaboud:2017phn,Aaboud:2017ejf,Aaboud:2018zjf,Aaboud:2017phn}. In
	contrast, color singlets are less conspicuous at a hadron collider, due to their electroweak-size production cross section, so that weaker constraints are expected. Nevertheless, several studies point to the exclusion of new-physics scenarios with typical masses ranging from $100$ to $500$~GeV \cite{Aaboud:2018sua,Sirunyan:2018ubx}. Yet, such limits commonly assume {\it optimized} scenarios -- characterized e.g.\ by the overwhelming dominance of specific channels in the decay chain -- which are not necessarily realized in ultraviolet-complete models. Therefore, it appears meaningful to assess the impact of these searches for a full (UV-complete) model.
	
	A popular family of models for physics beyond the SM is that of softly-broken supersymmetric (SUSY) extensions of the SM \cite{Nilles:1983ge,Haber:1984rc}. Originally motivated by the stabilization of the Higgs mass at the electroweak scale against radiative corrections from GUT/Planck-scale new-physics, this class of models also produces a dark-matter (DM) candidate in the presence of a (strict or approximate) R-parity: the lightest SUSY (R-odd) particle (LSP). Several arguments, such as the $\mu$-problem \cite{Kim:1983dt} or the naturalness of the Higgs mass, advocate the necessity to look beyond the Minimal SUSY Standard Model (MSSM), the minimal and by far most studied SUSY-inspired model. A departure from minimality also means the opening of potentially new effects in the phenomenology, which could e.g.\ complicate the reading of limits at colliders.
	
	In this paper, we will consider the Next-to-MSSM (NMSSM) \cite{Ellwanger:2009dp,Maniatis:2009re}, a simple singlet-extension of the MSSM. Beyond allowing for a solution to the $\mu$-problem, the NMSSM is often considered for the phenomenology of its extended Higgs sector -- consult e.g.\ \cite{Domingo:2015eea,Ellwanger:2015uaz,Conte:2016zjp,Guchait:2016pes,Badziak:2016tzl,Das:2016eob,Cao:2016uwt,Das:2017tob} for a few recent discussions and summaries. From the perspective of DM, the presence of an additional singlino component in the neutralino sector -- beyond the bino, wino and higgsinos of the MSSM --, opens up new scenarios satisfying the relic density \cite{Belanger:2005kh,Cerdeno:2007sn,Hugonie:2007vd}, possibly at low mass \cite{Vasquez:2010ru,Ellwanger:2014dfa,Han:2015zba}, even though the viability of this later option has been questioned \cite{Mou:2017sjf}.
	
	Another interesting aspect of the NMSSM phenomenology consists in the opening of new mechanisms at colliders. Refs.~\cite{Ellwanger:2014hia,Ellwanger:2014hca} have insisted on the impact that a light singlino LSP could have on squark / gluino searches -- see also \cite{Chakraborty:2015xia,Kim:2015dpa,Titterton:2018pba}. Here, we wish to focus on the collider searches applying to the neutralinos and charginos -- the superpartners of the Higgs and gauge bosons. Already at the level of the MSSM \cite{Aaboud:2016wna,Athron:2018vxy}, the multi-lepton searches obviously perform less efficiently than in the simplified framework in which they are presented. In the NMSSM, we can first stress the obvious difference with the MSSM due to the presence of the singlino component. The potential reach of LHC searches for this scenario was assessed in \cite{Xiang:2016ndq}. Ref.~\cite{Ellwanger:2016sur,Kim:2014noa,Ellwanger:2018zxt} recently pointed out the robustness of the higgsino-singlino DM scenario in view of lepton-signature searches at the LHC. In addition, the NMSSM superpotential opens up several couplings between the higgsino and the Higgs sectors and, in particular, couplings to the neutral singlet Higgs bosons. These features can have an impact on the properties of the final state at colliders, either through an alteration of the kinematic variables -- e.g.\ opening of compressed configurations due to the presence of an additional neutralino state and production of softer final states -- or through a weakened relevance of the standard channels -- e.g.\ electroweakino decays through Higgs bosons would replace light leptons in the final state by hadronic $\tau$'s, which are harder to identify experimentally. We will show how these characteristics may considerably affect the relevance of the multi-lepton searches performed by the ATLAS and CMS collaborations \cite{Aad:2015jqa,Aad:2014vma,Aad:2014nua,atlas-conf-2013-036,CMS-PAS-SUS-16-039,ATLAS-CONF-2017-039,Aaboud:2017nhr,CMS:2017fij,Aaboud:2017leg} in order to investigate the neutralino and chargino sectors at moderate masses. The prospects or constraints of the LHC searches on DM-inspired NMSSM scenarios have already been discussed in the past, see e.g.\ \cite{Xiang:2016ndq,Ellwanger:2016sur,Kim:2014noa,Ellwanger:2018zxt,Guo:2014gra,Cao:2014efa,Barducci:2015zna,Cao:2016nix,Cao:2018rix}. Here, however, while we still consider the upper bound on the DM relic density as a selection criterion, we wish to perform a more collider-oriented, less DM-prejudiced analysis, which we believe to be justified as several mechanisms could unsettle the identification of the LSP with today's DM.
	
	To this end, we carry out a scan on the NMSSM Higgs sector, using \texttt{NMSSMTools} \cite{Ellwanger:2004xm,Ellwanger:2005dv,NMSSMTools:website} and applying phenomenological limits of various origins encoded in this public tool; additionally, the DM relic density and the decays of SUSY particles are computed using \texttt{MicrOMEGAs} \cite{Belanger:2001fz,Belanger:2004yn,Belanger:2005kh,MicrOMEGAs:website} and \texttt{NMSDECAYS} \cite{Muhlleitner:2003vg,Das:2011dg}. Among the points satisfying all the constraints, we delimit several scenarios involving a neutralino--chargino sector at moderate masses (below $500$~GeV) and extract several points for testing against LHC SUSY searches: this last test is carried through \texttt{CheckMATE 2} \cite{Drees:2013wra,Dercks:2016npn,webpage,manager}, which is based on the fast detector simulator \texttt{Delphes} \cite{deFavereau:2013fsa}, after event-generation through \texttt{Pythia} \cite{Sjostrand:2014zea}. We then discuss how the NMSSM phenomenology impacts the collider searches in each scenario and suggest complementary signatures to help cover the parameter space. Our approach differs from that of the recent paper \cite{Cao:2018rix} in that we discuss the collider constraints on the the electroweakino sector independently of the DM detection observables, since the latter depend on additional (e.g.\ astrophysical) assumptions, and fine-tuning measures. We also target specific NMSSM scenarios, providing a less blind coverage of the NMSSM parameter space. Finally, we restrict to the scenario where all the scalar SUSY particles are comparatively heavy, hence testing the electroweakino sector as the only source of new-physics particles close to the electroweak scale (except for possible Higgs-singlet states).

	In the following section, we discuss our strategy for investigating the parameter space of the NMSSM and selecting scenarios involving the neutralino and chargino sectors. Then, we consider the impact of the multi-lepton searches at the LHC. We base this analysis both on general features of the scan and on the particular properties of specific test-points. Finally, we briefly discuss the expected reach of the High-Luminosity run at the level of the test-points that are still allowed after taking into account the early $13$~TeV result and suggest additional search strategies, before a short conclusion.
	
	\section{Investigating the neutralino and chargino sector of the NMSSM}
	
	\subsection{General considerations}
	In this work, we are considering the neutralino and chargino sectors of the NMSSM. For simplicity, we restrict ourselves to the CP-\ and $Z_3$-conserving NMSSM below. In this section, we remind a few general features relative to neutralinos and charginos in the NMSSM as well as their interactions. For completeness, we indicate the form of the superpotential below, though most of our notations follow \cite{Ellwanger:2009dp}:
	\begin{equation}\label{supo}
	W_{\mbox{\tiny NMSSM}}=\lambda \hat{S}\hat{H}_u\cdot\hat{H}_d+\frac{\kappa}{3}\hat{S}^3+W_{\mbox{\tiny Yukawa}}
	\end{equation}
	
	At first sight, the NMSSM neutralino and chargino sectors are very similar to their MSSM counterparts. The interactions of the bino and winos are fixed by the $SU(2)\times U(1)$ gauge symmetry. The soft SUSY-breaking Lagrangian provides the gaugino mass-terms $M_1$ and $M_2$. The mass of the doublet higgsinos originates in the effective $\mu$-term of the superpotential -- generated dynamically in the $Z_3$-conserving NMSSM when the gauge-singlet superfield takes its vacuum expectation value (vev): $\mu_{\mbox{\tiny eff}}=\lambda s$, $s\equiv\left<\!\right.\hat{S}\left.\!\right>$. The gaugino-higgsino mixing is generated by the SUSY-gauge interactions when the electroweak symmetry is broken. Beyond the MSSM, however, the NMSSM includes one additional
	fermionic component, singlet under electroweak interactions: the singlino. The superpotential determines the singlino-singlet interactions, hence the singlino mass $2\kappa s$, as well as the mixing of the singlino with the doublet higgsinos, $\propto\lambda v$ (with $v\equiv(2\sqrt{2}G_F)^{-1/2}$ the electroweak-breaking vev, and $G_F$ the Fermi constant).
	
	These considerations may be summarized in writing down the mass matrices of the charginos and neutralinos. In the interaction bases of the charged gauginos and higgsinos, $\psi^+=(-\imath\lambda^+,\psi_u^+)^T$, $\psi^-=(-\imath\lambda^-,\psi_d^-)$, the chargino mass-term reads:
	\begin{equation}
	{\cal L}\ni-\psi^-{\cal M}_{\chi^{\pm}}\psi^++h.c.\hspace{0.5cm},\hspace{0.5cm}{\cal M}_{\chi^{\pm}}=\begin{pmatrix}
	M_2 & g_2v_u\\
	g_2v_d & \mu_{\mbox{\tiny eff}}
	\end{pmatrix}
	\end{equation}
	For the neutralinos in the base $\psi^0=(-\imath\lambda_1,-\imath\lambda^3_2,\psi_d^0,\psi_u^0,\psi_S)^T$, the mass-term reads:
	
	\begin{equation}
	{\cal L}\ni-\frac{1}{2}\psi^{0\,T}{\cal M}_{\chi^{0}}\psi^0+h.c.\hspace{0.5cm},\hspace{0.5cm}{\cal M}_{\chi^{0}}=\begin{pmatrix}
	M_1 & 0 & -\frac{g_1v_d}{\sqrt{2}} & \frac{g_1v_u}{\sqrt{2}} & 0 \\
	0 & M_2 & \frac{g_2v_d}{\sqrt{2}} & -\frac{g_2v_u}{\sqrt{2}} & 0 \\
	-\frac{g_1v_d}{\sqrt{2}} & \frac{g_2v_d}{\sqrt{2}} & 0 & -\mu_{\mbox{\tiny eff}} & -\lambda v_u \\
	\frac{g_1v_u}{\sqrt{2}} & -\frac{g_2v_u}{\sqrt{2}} & -\mu_{\mbox{\tiny eff}} & 0 & -\lambda v_d \\
	0 & 0 & -\lambda v_u & -\lambda v_d & 2\kappa s
	\end{pmatrix}
	\end{equation}
	Diagonalizing these mass-matrices, one obtains the chargino mass-states (at tree-level), $\chi^-_i=U_{ij}\psi^-_j$ and $\chi^+_i=V_{ij}\psi^+_j$, as well as the neutralino mass-states $\chi^0_i=N_{ij}\psi^0_j$, where $U$, $V$ and $N$ are orthogonal mixing-matrices.
	
	Beyond the extended neutralino sector, the interactions of the higgsinos in the NMSSM differ from those of their MSSM counterparts. Indeed, the superpotential produces higgsino / singlino couplings to the Higgs sector, involving singlet as well as doublet Higgs components. This has the important phenomenological consequence that neutralino decays and production-channels in the NMSSM may more easily employ a Higgs mediator. Such an affinity to Higgs bosons affects the efficiency of searches through leptonic final states, due to the suppressed Higgs couplings to light leptons. For a pure singlino state, the Higgs sector is actually the only point of contact with SM matter. Mass-mixing with the higgsino components (and secondarily with gauginos) may generate direct couplings to gauge-bosons for a mostly-singlino state however. Depending on the configuration of the spectrum, the 	phenomenology of such a state could be dominated by its subdominant higgsino-gaugino components or by its naive singlino-like couplings. 
	
	This discussion shows that the Higgs sector could play a significant part in the phenomenology of NMSSM neutralinos and charginos. Beyond the two CP-even and the unique CP-odd neutral doublet states in the MSSM, the NMSSM involves one additional CP-even and one CP-odd singlet components. Singlet-doublet mixing appears at tree-level. It tends to dominate the couplings of mostly-singlet states to SM matter, since pure singlet components only interact with the Higgs and higgsino sectors otherwise. The production cross section of singlet states at colliders is thus suppressed, opening the path to realistic scenarios involving singlets lighter than $125$~GeV and possibly as light as a few GeV -- in this latter case, however, severe constraints from flavor physics or the non-observation of sizable unconventional decays of the Higgs-state at $\sim125$~GeV must be taken into account, see e.g.\ \cite{Domingo:2015eea}. In contrast, singlet states directly couple to higgsino and singlino components via the superpotential parameters $\lambda$ and $\kappa$ -- see Eq.(\ref{supo}) -- leading to a possible impact on the phenomenology of neutralinos and charginos when $\lambda$ and $\kappa$ are of order $0.1 - 1$ (i.e.\ far from the MSSM limit $\lambda\sim\kappa\to0$). For example, the singlet Higgs states open new Higgs funnels for the annihilation of the LSP in the early Universe \cite{Belanger:2005kh,Cerdeno:2007sn,Hugonie:2007vd}. In particular, light singlets allow for realistic light DM scenarios \cite{Vasquez:2010ru,Ellwanger:2014dfa,Han:2015zba}. Additionally singlet states could enter neutralino decay chains, typically leading to $b\bar{b}$ or $\tau^+\tau^-$ signatures: their couplings to higgsino and singlino components may supersede gauge couplings and, if light, they may easily be exchanged on shell.
	
	Therefore, despite the apparent closeness between the MSSM and NMSSM neutralino and chargino sectors, we can expect sizable differences in the phenomenology of both models at colliders, which we aim at investigating in the following sections.

	\subsection{Generating the spectra}
	We explore the parameter space of the CP- and $Z_3$-conserving NMSSM with the public spectrum generator \verb|NMSSMTools_5.1.0| \cite{Ellwanger:2004xm,Ellwanger:2005dv,NMSSMTools:website}. This tool includes leading radiative corrections to the masses and couplings of the Higgs and SUSY particles. Higgs decays are also calculated in this package through an extension of \texttt{HDECAY} \cite{Djouadi:1997yw,Djouadi:2018xqq} to the NMSSM. In the case of light -- generally singlet-dominated -- states, \texttt{NMSSMTools} now employs the more consistent description outlined in \cite{Narison:1989da,Domingo:2011rn,Domingo:2016yih}. Similarly, \texttt{NMSDECAYS} \cite{Das:2011dg} -- generalizing \texttt{SDECAY} \cite{Muhlleitner:2003vg} to the NMSSM -- computes the decay widths and branching ratios of the SUSY particles.
	
	We perform a random scan over a region of the parameter space characterized by the following input. $\lambda\in[0.001,0.7]$, $\kappa \in [-0.7,0.7]$, $\tan\beta \in[1,30]$, $\mu_{\mbox{\tiny eff}} \in[-1,1]$~TeV, $M_1 \in[-1,1]$~TeV, $M_2 \in[0.01,1]$~TeV.	We fix the mass of the heavy doublet Higgs states via the input $M_A=1$~TeV -- here, $M_A$, the diagonal doublet-mass entry in the CP-odd mass-matrix, substitutes the trilinear soft coupling $A_{\lambda}$ and largely determines the mass of the CP-even, CP-odd and charged doublet Higgs states -- but we scan over the mass of the singlet states via the condition $m_P \in[1,1000]$~GeV -- $m_P$ represents the diagonal singlet-mass entry in the CP-odd mass-matrix and replaces the trilinear soft coupling $A_{\kappa}$. We are indeed chiefly	interested in the impact of Higgs states beyond the MSSM. Concerning the SUSY scalar sector, since our focus is that of moderately light neutralinos / charginos, we choose rather heavy scales for the masses of squarks and sleptons, beyond the naive reach of Run-1 searches: the slepton masses are fixed at $1$~TeV while the squark masses are varied between $2$ and $15$~TeV -- this wide range is motivated by the 	condition on the mass of the SM-like Higgs boson, which we cannot set to $\sim125$~GeV as an input, but should reach this value (within theoretical and experimental uncertainties) when the squark scale scans over the interval $[2,15]$~TeV. The trilinear sfermion couplings $A_f$ are in the range $[-2,2]$~TeV. Finally, $M_3=3$~TeV should place the gluinos at a relatively safe scale in view of current limits.
	
	This choice of input is criticizable in many ways. The upper bound $|\lambda|, |\kappa|<0.7$ is the typical limit set by the condition of perturbativity of the couplings up to the Grand-Unification (GUT) scale. Very large values of $\tan\beta>30$ generally result in sizable enhancements of the heavy Higgs couplings, which can lead to tensions in the flavor sector or in direct searches. Concerning the gaugino and higgsino masses, we are mostly interested in light states, since the electroweak-size production cross section typically falls out-of-reach
	of the LHC sensitivity for large mass-suppression, motivating our upper limit of $1$~TeV. Regarding the Higgs masses, our decision of fixing $M_A=1$~TeV excludes the mediation of DM annihilation by a heavy doublet state, except when the LSP has a mass of $\sim500$~GeV (on the fringe of the relevant range for collider multi-lepton searches). This scenario is not our focus, however, as we are interested in effects beyond the MSSM. Nevertheless,
	$M_A=1$~TeV also tends to suppress the singlet-doublet mixing among Higgs states, which impacts the affinity of light singlet states to SM-like particles. The relevance of LHC colored searches for particles below the TeV range justifies our choice of restricting to heavy squark states. In the case of the sleptons, however, this decision is less motivated as these are weakly-interacting particles. Sleptons are typical t-channel mediators for interactions between the chargino / neutralino sector and SM particles. Light smuons are also motivated by the excess in the measurement of the anomalous magnetic moment of the muon \cite{Bennett:2006fi}. Again, such configurations already occur in the MSSM, which explains our decision of discarding them in the current discussion.
	
	\subsection{Phenomenological limits on the scan}\label{subsec:pheno_constraints}
	\texttt{NMSSMTools} is equipped with several tests allowing for the selection of points of reasonable phenomenological relevance. Here, we provide a brief summary of the constraints that we choose to apply.
	\begin{itemize}
		\item A first class of limits results from general considerations on the perturbativity of the couplings up to the GUT scale, the naturalness of soft SUSY-breaking mass terms, the stability of the spectrum (absence of tachyonic masses) or the potential.
		\item \texttt{NMSSMTools} applies limits on the NMSSM Higgs sector originating in direct searches at LEP \cite{Barate:2003sz,Schael:2010aw}, the TeVatron \cite{Abazov:2009wy,Aaltonen:2009ke,Abazov:2009aa} or the LHC. In the latter case, the properties of the observed Higgs state are tested in a global fit 
		\cite{Belanger:2013xza} while several constraints from unsuccessful searches \cite{ATLAS:2011pka,CMS:2013hja,ATLAS:2014vga,CMS:2014hja,CMS:2015iga,
			CMS:2016cel,CMS:2016cqw,Aad:2015oqa,Khachatryan:2015wka,Khachatryan:2015nba} -- in particular searches for Higgs-to-Higgs decays involving a light singlet state -- are also considered.
		\item Additionally, the scenario involving a light (singlet-dominated) Higgs state is sensitive to constraints from bottomonium decays and spectroscopy, which are implemented according to 		\cite{Domingo:2008rr,Domingo:2009tb,Domingo:2010am}.
		\item Limits on the invisible $Z$-decays are applied following the SM estimate of \cite{Freitas:2014hra}.  
                  
		\item SUSY-searches at LEP \cite{LEPSUSYsearches} are included in the form of cuts on the SUSY masses as well as limit on stop and sbottom decays.
		\item The flavor-observables implemented in \texttt{NMSSMTools} have been described in \cite{Domingo:2007dx,Domingo:2015wyn}.
	\end{itemize}
        In addition, bounds from invisible SM Higgs decays are implemented, which can constrain the parameter space of the electroweakino sector \cite{Dreiner:2012ex}.
	Thus a sizable collection of phenomenological limits is employed, with the notable exception of LHC SUSY searches. Furthermore, since we decided to freeze the slepton mass at $1$~TeV, we also choose to discard limits on the anomalous magnetic moment of the muon included in \texttt{NMSSMTools} \cite{Domingo:2008bb}.
	
	DM observables can be computed via an interface with \texttt{micrOMEGAs} \cite{Belanger:2005kh}. A strong assumption behind the application of corresponding limits is that the LSP of the NMSSM is the actual DM of the Universe and that it is thermally produced. We note that there is no deep reason in making this identification, as other production modes, other sources of DM or decays of the LSP (of the NMSSM) into e.g.\ a lighter gravitino or through small R-parity-violating terms could be invoked. Yet, we choose to consider the measured DM relic density \cite{Hinshaw:2012aka,Ade:2013zuv} as an upper bound on thermal LSP relic production in the early Universe. Limits from direct detection searches \cite{Tan:2016zwf,Akerib:2016vxi,Aprile:2017iyp,Aprile:2018dbl,Amole:2016pye,Akerib:2016lao,Fu:2016ega}
	depend on further astrophysical assumptions as to the distribution of DM in our galaxy, which could be questioned further if our LSP only represents a fraction of the DM relics. The complementarity of these searches is frequently invoked and one could derive the associated limits under e.g.\ the simple assumption that the limits from direct searches can be rescaled in proportion to the amount that our LSP relics represent with respect to the measured DM relic density. However, for simplicity, we will not consider them, as their impact is comparatively orthogonal to that of collider searches.
	
	\subsection{NMSSM electroweakino scenarios}\label{subsec:scenarios}
	After performing the random scan and applying the phenomenological limits described in the previous subsections, we obtain a large number of viable candidate spectra in the NMSSM. While we wish to investigate limits from LHC SUSY searches, we can only perform these tests over a limited number of spectra, due to the large amount of computer resources needed in order to simulate the events. Therefore, we choose to restrict this collider analysis to a few thousand points that we classify into specific configurations of the chargino / neutralino / Higgs spectra. Although this restriction means that we could be testing too few points to get a fully representative sample of the NMSSM electroweakino phenomenology, we try to compensate this feature by targeting NMSSM-specific effects. In view of the typical range of neutralino and chargino masses of the MSSM for which multi-lepton searches are relevant (see Fig.2 of \cite{Aaboud:2016wna} for Run-1 and Fig.14-18 of \cite{Sirunyan:2017lae} for Run-2), we focus on NMSSM spectra with LSP mass below $200$~GeV and NLSP mass below $500$~GeV. In fact we even maximize the numerical effort on points with LSP masses below $100$~GeV and NLSP masses below $300$~GeV. The characteristics of each scenario are described below.
	
	\paragraph{MSSM-like spectra --}
	While the MSSM-limit of the NMSSM is characterized by $\lambda\sim\kappa\ll1$, we do not expect sizable effects beyond the MSSM in the neutralino / chargino searches if none of the singlet and singlino states intervenes in the production and decays of the lighter neutralino / chargino states (with mass below $500$~GeV). By extension, if all singlino and singlet states are heavy (beyond $\sim500$~GeV), the outcome of collider searches should be comparable to that obtained in the MSSM case. We thus define a first scenario where singlino and singlet states appear with a mass beyond $500$~GeV. This will serve as a control region for comparison with the MSSM results. This sample contains $\sim3000$ points. These cluster in the $Z$-\ and Higgs-funnel as well as in the chargino-coannihilation region due to the limits associated with the DM relic density. The LSP is then dominantly bino, with possible winos or higgsinos in the upper mass-range ($m_{\text{LSP}}\gsim100$~GeV).
	
	\paragraph{Singlino LSP --}
	The second scenario involves a singlino-dominated LSP (with mass below $200$~GeV). This sample contains $\sim3500$ spectra. LSP masses go down to a few GeV, where DM annihilation typically proceeds via a singlet-Higgs funnel. Beyond the low-mass region, this annihilation channel may intervene for any choice of LSP / NLSP masses, so that the structures of $Z$/Higgs-funnels and chargino-coannihilation region are blurred. In the $Z_3$-conserving NMSSM, however, a sum rule among the masses makes it difficult to reach NLSP masses beyond a few hundred GeV: comparatively light higgsino states are indeed expected. In the presence of a singlino LSP, all the SUSY decay chains end up with this genuine NMSSM state, which leads to an obvious alteration of the phenomenology as compared to the MSSM setup.
	
	\paragraph{Singlino NLSP --}
	The impact of a singlino NLSP on SUSY searches is less obvious than that of a singlino LSP. Indeed, this NLSP state is typically harder to produce in collisions of SM particles than the heavier gauginos and higgsinos. In addition, the decay chains of the heavier states could be blind to the presence of this NLSP. In such a case, the existence of a light singlino NLSP is largely devoid of phenomenological consequences. Therefore, we restrict this scenario further to the case where the singlino intervenes at more than $30\%$ in the decay of heavier neutralino / chargino states. We keep a total of $\sim2500$ points in this scenario. As could be expected, the structures of Higgs/$Z$-funnels (determined by the constraint on the DM relic density) emerge again, as in the MSSM-scenario. On the other hand, the condition of a singlino NLSP is largely incompatible with the coannihilation region. The presence of a singlino NLSP in the SUSY decay chains could lead to various effects. In particular, it adds a new ladder in the decay chain, which possibly increases the number of SM-particles in the final state and tends to make them softer (the spectrum is more compressed).
	
	\paragraph{Decays into Higgs singlets --}
	The presence of light Higgs singlet states can affect the phenomenology of neutralinos and charginos, as the SUSY particles may now sizably decay toward such a Higgs final state. This impacts the multiplicity of leptons in the final state, since only the less-efficiently detected $\tau$'s are substantially produced in Higgs decays. We define a sample of points where the singlet Higgs intervenes at more than $10\%$ in the decays of the NLSP or a close-by neutralino / chargino state (with mass within $10$~GeV of the NLSP). Our scan contains $\sim1250$ such points, Most of the time, the CP-odd state is lightest (with $\mathcal{O}(100)$ exceptions), but both CP-even and CP-odd states are often simultaneously light. We note that this scenario has a non-vanishing overlap with the light singlino scenario. Yet, we checked that only $\mathcal{O}(100)$ out of our $\sim1250$ points involve a singlino LSP or NLSP. In fact, most of the spectra contain a bino-like LSP with mass close to half of the $Z$ or the SM-Higgs masses.
	
	\paragraph{Higgs singlet on LSP annihilation threshold --}
	Our final sample consists of points where the LSP annihilation is mediated by a Higgs singlet state. The spectra that we retain here satisfy the approximate condition $m_S\simeq 2\, m_{\mbox{\tiny LSP}}$, where $m_S$ is the mass of either the CP-even or the CP-odd singlet state and $m_{\mbox{\tiny LSP}}$ is the mass of the LSP. Additionally, we exclude the traditional $Z$/Higgs-funnel and chargino-coannihilation region based on the mass-contours that we obtained in the MSSM-like scenario. We note that we already encountered singlet-mediated annihilation in the context of singlino LSP. However, the large majority of the points satisfying the previous conditions in our scan involve a bino-LSP (occasionally a higgsino). Our sample gathers $\sim3000$ of such points.
	
	\subsection{Collider test -- multi-lepton signatures}
	\paragraph{Collider signatures --}
	Before discussing the relevant SUSY searches, we briefly summarize the important collider signatures. Since the colored superpartners as well as electroweak sfermions have very small production cross sections (due to their mass) in the spectra under investigation, only neutralinos, $\chi_i^0$, and charginos, $\chi_m^\pm$, are kinematically accessible at the LHC,
	\begin{equation}\label{eq:production}
	pp\rightarrow\chi_i^0\chi_j^0,\quad pp\rightarrow\chi_i^0\chi_m^\pm,\quad pp\rightarrow\chi_m^\pm\chi_n^\mp, 
	\end{equation}
	where $i,j=1,\dots,5$ and $m,n=1,2$. Here, we have omitted single electroweakino production since the production cross section is negligible for decoupled squarks. A typical cross section of chargino pair production in $p-p$ collisions with $8$~TeV center-of-mass energy (resp.~$13$~TeV) can be as large as $\mathcal{O}(2)$ pb (resp.\ $\mathcal{O}(4)$ pb) for charginos at $100$~GeV while chargino--neutralino production may yield a production cross section of $\mathcal{O}(5)$~pb (resp.\ $\mathcal{O}(10)$~pb). The decay chains are relatively complicated and mainly depend on the composition of the electroweakinos. If kinematically allowed, the dominant decay modes involve on-shell weak-gauge or Higgs bosons:
	\begin{subequations}\begin{align}
		&\chi_i^0\rightarrow W^\pm\chi_m^\mp,& &\chi_i^0\rightarrow H^\pm\chi_m^\mp,\\
		&\chi_i^0\rightarrow Z\chi_j^0,& &\chi_i^0\rightarrow \Phi\chi_j^0,\\
		&\chi_i^\mp\rightarrow Z\chi_j^\mp,& &\chi_i^\mp\rightarrow \Phi\chi_j^\mp\\
		&\chi_i^\mp\rightarrow W^\mp\chi_j^0,& &\chi_i^\mp\rightarrow H^\mp\chi_j^0,
		\end{align}\end{subequations}
	where $\Phi=H_1,H_2,H_3,A_1,A_2$. If such channels are kinematically inaccessible, the neutralinos and charginos decay into three body final states via off shell Higgs  and gauge bosons,
	\begin{subequations}\begin{align}
		\chi_i^0&\rightarrow \chi_j^0 f\bar f,\\
		\chi_i^0&\rightarrow \chi_m^\pm f\bar f^\prime,\\
		\chi_m^\pm&\rightarrow \chi_i^0 f\bar f^\prime,
		\end{align}\end{subequations}
	where $f$, $f^\prime$ denote SM fermions. If the mass-splitting between electroweakino states is below a few GeV, the partons in the final state cannot be treated as free particles so that the explicit decays into pions must be taken into account, e.g.\ $\chi_1^+\rightarrow \pi^+ \chi^0_1$ \cite{Chen:1996ap}. In addition, when the phase-space is extremely reduced and the electroweakino states have a sizable higgsino nature, the loop-induced channel 
	\begin{equation}
	\chi_2^0\rightarrow \chi_1^0\gamma,
	\end{equation}
	can acquire a sizable branching ratio or even dominate the $\chi_2^0$ decays for moderate values of $\tan\beta$ \cite{Haber:1988px}.
	
	\paragraph{Experimental searches --}
	In general, leptonic final states benefit from a much stronger sensitivity at colliders, as compared to hadronic final states, due to their very clean signature and the reduced SM backgrounds. As a consequence experimental searches for electroweakinos mostly consider final states with high-$p_T$ leptons and missing transverse momentum. In experimental analyses, leptons are divided into two classes, the so called light leptons $e$ and $\mu$ and the hadronically decaying $\tau$'s. Throughout this work we demand that the decays are prompt, so that searches for long-lived particles, such as disappearing track analyses \cite{Aaboud:2017mpt}, are not relevant.
	
	{\it Run-1:}
	The lepton plus SM Higgs search targets associated chargino--neutralino pair production with subsequent decay into light leptons, large missing transverse momentum and a Higgs boson \cite{Aad:2015jqa}. The Higgs boson is reconstructed in the $b\bar b$ (SR$\ell b b$), $\gamma\gamma$ (SR$\ell \gamma \gamma$) and $WW$ (SR$\ell\ell$) final state -- in the latter case a second lepton (expected from the $W$ boson decay) with the same sign (SS) charge as the first lepton (expected from the chargino decay) is requested.
	
	The dilepton and large missing transverse momentum search aims at discovering electroweak production of charginos and neutralinos as well as production of slepton pairs \cite{Aad:2014vma}. There, the authors demand exactly two leptons (electrons, muons) and large missing transverse momentum. For chargino pair production with intermediate sleptons, they impose a strict cut on the transverse mass $m_{T2}$ and the events are categorized in same flavour (SF) and different flavour (DF) classes, namely the following final states $e^+e^-$, $\mu^+\mu^-$, $e^+\mu^-$. Chargino pair production followed by decays into $W$'s and the LSP are targeted by three signal regions (SR-$WW$). 
	A parent $Z$ boson is also consistent with the signal regions including $e^+e^-$ or $\mu^+\mu^-$ in the final state. In this case, additional jet activity is required by the search. This search targets production processes such as $pp\rightarrow\chi_1^\pm \chi_2^0$ with a hadronically decaying $W$ boson (SR-Zjets). We have not considered the complementary search targeting two $\tau$'s plus large missing transverse momentum and a jet veto since it has not been implemented into {\tt CheckMATE}.
	
	The trilepton study searches for direct production of chargino--neutralino pair, further decaying into three leptons (electron, muon, tau) in association with large missing transverse momentum \cite{Aad:2014nua}. Here, SM-Higgs as well as $Z$-boson mediated $\chi_2^0$ decays are taken into account. In the MSSM, this search usually provides the highest sensitivity to a light electroweakino sector. Three classes of signatures are considered in the final state: three light leptons, i.e.\ $e$ or $\mu$ (SR0$\tau$), two light leptons and a hadronically decaying $\tau$ (SR1$\tau$), finally a single lepton plus two $\tau$'s (SR2$\tau$). These are further split into 	24 signal regions targeting chargino--neutralino pair production with subsequent decays into sleptons, gauge bosons or the SM Higgs boson which translate into `same flavour opposite sign' (SFOS) requirement or veto, missing transverse cut, invariant mass cuts and kinematic cuts on opposite sign (OS) lepton pairs. 
	
	Finally, the four lepton analysis targets events with four or more leptons \cite{atlas-conf-2013-036}. The signal regions demand at least four light leptons (SR0Z) or exactly three light leptons and a hadronically decaying tau (SR1Z) requiring $Z$ candidates. This search is motivated by higgsino-like $\chi_2^0\chi_3^0$ pair production where both higgsinos decay into a $Z$ and the LSP . Moreover, it includes signal regions targeting R-parity violating decay modes, slepton-induced decays and decays via gauge mediated SUSY breaking scenarios.
	
	{\it Run-2:} 
	The ATLAS and CMS collaboration have published a plethora of new searches in Run-2. The CMS search \cite{CMS-PAS-SUS-16-039} for multilepton final states looks for charginos and neutralinos in signatures with either two same-sign light leptons or with three or more leptons, while allowing up to two hadronically decaying $\tau$'s and demanding little hadronic activity as well as missing transverse momentum. All final state topologies are covered by a large number of signal regions. E.g.\ the dilepton phase space is split into multiple signal regions characterized by initial-state-radiation veto, missing transverse momentum and the transverse momentum of the dilepton pair. The `three or more lepton' categories are classified according to the number of light leptons and hadronic taus. The signal regions with label A have at least one light OSSF pair among the three light leptons and are categorised according to missing transverse momentum, transverse mass, and invariant dilepton mass. Signal regions of type B do not contain an OSSF pair. Events with three leptons and at least one hadronic tau which contain an OSSF pair are further tested with $m_{T2}$ as a discriminating variable and the corresponding signal region are in category C. If no OSSF is found the events are split either in events with OS or SS lepton pairs corresponding to categories  D and E, respectively. The last category F contains events with two hadronic taus. Finally, events with more than three leptons are classified according to the number of OSSF pairs and number of hadronic taus and intervals of missing transverse momentum and are labelled with categories G-K.
	
	Likewise, ATLAS presented an electroweakino search in two and three lepton final states \cite{ATLAS-CONF-2017-039}. They divide their study into three search strategies. The first looks for two leptons and demands a jet veto targeting chargino pair production. This final state channel is further binned according to the transverse mass and the dilepton invariant mass, while the signal regions are split between SF and DF dilepton pairs (SR2-SF, SR2-DF). The second search strategy focuses on dileptons and additional hadronic activity which is optimized for the associated chargino-neutralino pair production with subsequent decays into gauge bosons, where the $W$ decays into two jets and the dilepton pair originates from the $Z$ decay. The signal regions are organized according to the size of the mass-splitting between the NLSP and the LSP and is denoted with SR-low, SR2-int and SR-high. The final search strategy targets chargino-neutralino production leading to the trilepton and missing transverse momentum final state. The trilepton final state signatures are binned in missing transverse momentum, transverse mass and the transverse momentum of the least energetic lepton denoted by SR3-WZ-0J. If a $b$-jet veto is required the signal region identifier is SR3-WZ-1J. In all searches, the leading lepton is required to have $p_T^{\rm min}>25$~GeV.
	
	Ref.~\cite{Aaboud:2017nhr} searches for two hadronically decaying taus and missing transverse momentum. The search targets electroweakino production modes with decays via intermediate third generation sleptons. The selection cuts are relatively generic so that the search might be sensitive to our scenarios. The search demands an OS $\tau$ pair with moderate requirements on the $\tau$ transverse momentum and missing transverse momentum. To further discriminate the signal from the background, the transverse mass $m_{T2}$ is exploited in order to remove $t \bar t$ and $WW$ events. 
	
	In Run-2 the experimental collaborations have started to probe the compressed electroweakino sector with soft leptons in the final state which is an extremely challenging signature at hadron colliders. In Ref.~\cite{CMS:2017fij} the authors probe two low-momentum OS leptons and missing transverse momentum. They target compressed charginos and neutralinos decaying via off shell SM gauge bosons into the LSP. With simplified model assumptions they are able to exclude chargino masses up to $165$~GeV with mass differences of $7.5$~GeV between the NLSP and the LSP. They reach such sensitivity to compressed spectra because the signal leptons are only required to have $p_T^{\rm min}>5$~GeV. In addition, some moderate cuts on missing transverse momentum, invariant mass and the scalar sum of jet transverse momenta are imposed.
	
	ATLAS also presented a search \cite{Aaboud:2017leg} focusing on scenarios with compressed mass spectra. They could exclude higgsino (wino) scenarios with mass-splittings down to 2.5 (2.0) GeV. Signal electrons (muons) are requested to have $p_T^{\rm min}>$ 4.5 (4) GeV. Further cuts on missing transverse momentum, transverse mass and on ISR are demanded. Finally, the SR are binned in exclusive as well as inclusive bins of the invariant dilepton mass.
	
	Ref.~\cite{Cao:2018rix} also considered two additional CMS searches, from Refs.~\cite{Sirunyan:2017zss,Sirunyan:2017qaj}. Unfortunately, these searches had not been implemented into \texttt{Checkmate} at the time of our numerical scan, and could not be included a posteriori without a substantial numerical effort. However, we only expect a limited impact on our parameter points. \cite{Sirunyan:2017zss} focuses on the electroweak production of charginos and neutralinos leading to $WH$ events and seems to perform better than the corresponding search in \cite{CMS-PAS-SUS-16-039}. It selects events with an isolated lepton (from the $W$) and a $b\bar{b}$ pair (from a SM Higgs boson), applying cuts on the transverse as well as contranverse masses and considering two separate bins of missing transverse momentum. As a SM-like Higgs boson is explicitly targeted, we do not expect additional coverage in the scenarios with light singlet Higgs. The search of ref.~\cite{Sirunyan:2017qaj} targets final states with two opposite-charge, same-flavor leptons, jets and missing transverse momentum. The relevant topologies are strongly-produced electroweakinos, which subsequently employ a decay mode characteristic of a slepton-edge scenario. Such signals are certainly irrelevant for our benchmark points where we have assumed the sfermions to be heavy. However, this CMS search also considers $\chi_1^0$ pair production decaying into $ZZ$, $ZH$ and a light gravitino as well as $\chi_2^0\chi_1^\pm$ production with the characteristic $WZ$ final state. The relevant signal region is the on-Z search region, which is sensitive to hadronically decaying $W$- or $Z$-boson in association with the leptonically decaying $Z$-boson. The $WZ$ signal regions appears to perform slightly better than the corresponding search in \cite{CMS-PAS-SUS-16-039}, since the observed reach slightly improves on the expected one.
	
	\paragraph{Numerical Procedure --}\label{sec:num_signatures}
	Now we briefly discuss the numerical tools employed for the collider test. The searches discussed in the 
	previous paragraph have been implemented by the {\tt CheckMATE} collaboration \cite{Drees:2013wra,Dercks:2016npn,webpage,manager}. {\tt CheckMATE 2.0.26} is based on the modified fast detector simulation {\tt Delphes 3.4.1} \cite{deFavereau:2013fsa}. {\tt CheckMATE} tests whether a model point is 	excluded by comparing its expected signatures with all implemented experimental searches at the LHC. {\tt SLHA2} spectrum files are accepted as input and Monte Carlo (MC) events are generated with {\tt Pythia 8.223}.
	
	In Table~\ref{tab:lhc_searches} we list the searches implemented in {\tt CheckMATE} that are relevant for our 	analysis. The column on the left displays the arXiv number or the conference proceedings reference of the 	corresponding search. The second column shows the final state signature and the third column gives the total integrated luminosity. All studies listed in Table~\ref{tab:lhc_searches} have been validated against the results published by the experimental collaborations. Details on the validation can be found in the {\tt CheckMATE} manual and web page \cite{Drees:2013wra,webpage}.
	
	\begin{table}
		\begin{center}
			\begin{tabular}{c|l|l|c}
				$\sqrt{s} $ & Reference & Final State & $\mathcal{L}$ [fb$^{-1}$]\\
				\hline
				\hline
				8 TeV & 1501.07110 (ATLAS) \cite{Aad:2015jqa} & 1$\ell$+1$h$+\met & 20.1 \\
				& 1403.5294 (ATLAS) \cite{Aad:2014vma} & 2$\ell$+(jets)+\met & 20.1 \\
				& 1402.7029 (ATLAS) \cite{Aad:2014nua} & 3$\ell$+\met & 20.1 \\
				&ATLAS-CONF-2013-036 \cite{atlas-conf-2013-036} & 4$\ell$+\met & 20.1 \\
				\hline
				\hline
				13 TeV & CMS-PAS-SUS-16-039 \cite{CMS-PAS-SUS-16-039} & $\ge3\ell$+\met & 35.9\\
				& ATLAS-CONF-2017-039 \cite{ATLAS-CONF-2017-039} & $2(3)\ell$+\met& 36.1 \\
				& 1708.07875 (ATLAS)\cite{Aaboud:2017nhr} & 2$\tau_h$+\met & 36.1\\
				& CMS-PAS-SUS-16-048\cite{CMS:2017fij} & soft dilepton + \met & 35.9\\
				& 1712.08119 (ATLAS)\cite{Aaboud:2017leg} & soft dilepton + \met & 36.1\\
			\end{tabular}
		\end{center}
		\caption{Multi-lepton searches included in the collider test. The first column shows the center of mass energy. The second column provides the arXiv number or the conference proceedings identifier. The middle column denotes the final state which is targeted by the analysis and the last column displays the total integrated luminosity. $\ell$ denotes electron, muon as well as hadronic taus $\tau_h$.}
		\label{tab:lhc_searches} 
	\end{table}

	In order to estimate the efficiencies for all signal regions of all employed searches, hence the number of signal events for all searches, we first generate truth level MC events for each benchmark point with {\tt Pythia 8.223} \cite{Sjostrand:2014zea}. Here, we consider all processes summarized in Eq.~(\ref{eq:production}). The MC event generation for the simulation of the production of supersymmetric partners of the electroweak SM bosons is computationally quite expensive. The reasons are manifold such as relatively large production cross sections, the small leptonic branching ratios and the small efficiency of the signal regions. In general, this requires a large MC event sample. However, due to computational limits the maximally allowed number of MC events was set to 150000 events. As a result of this computational cost, we could not sample an {\it arbitrary} number of model points, as we discussed before. The corresponding total cross sections are computed at tree level and scaled up by a constant $k$-factor of $1.3$.
	
	For particle spectra involving relatively small mass-splittings with respect to the LSP, the simulation of additional radiation is important \cite{Dreiner:2012gx,Dreiner:2012sh}. However, we have not matched the partons from the exact matrix element calculation with the parton shower \cite{Alwall:2007fs} and thus our results in the compressed region have a large uncertainty. Finally, we removed the benchmark points for which {\tt Pythia} was not able to process the hadronization of the final states, due to very little available phase space.
	
	The truth level MC events together with the production cross sections are then passed on to {\tt CheckMATE}.
	Each model point is tested against all the analyses shown in Table~\ref{tab:lhc_searches}. {\tt CheckMATE} determines the optimal signal region among all the analyses with the largest expected exclusion limit. For this signal region, {\tt CheckMATE} compares the simulated signal with the actual experimental observation and determines whether the model point is excluded at the $95\%$ C.L.\ \cite{Read:2002hq} with the help of the following ratio, 
	\begin{equation} \label{r}
	r \equiv \frac{S-1.64\Delta S} {S_{\rm exp.}^{95}}\,, 
	\end{equation}
	where $S$, $\Delta S$ and $S_{\rm exp.}^{95}$ denote the number of signal events, the MC error and the experimentally determined 95$\%$ confidence level limit on $S$. The error due to the finite MC sample is $\Delta S = \sqrt{S}$. Here, we do not include systematic errors in the calculation of the ratio $r$ such as the theoretical uncertainty on the partonic production of electroweakinos and SUSY decay chains (higher-order, parametric, etc.), parton distribution, parton shower and luminosity uncertainties.
	
	We do not statistically combine signal regions since the correlations among them are not publicly available in 
	general. The exact value of $r$ delimiting the $95\%$ C.L.\ exclusion contour is a matter of discussion. 	We will consider a point as being clearly in tension with the experimental data if $r\ge 1.5$., i.e.\ if for the most sensitive signal region, the predicted number of signal events is by a factor of $1.5$ larger than the $95\%$ C.L. upper bound. If $r < 0.67$, the point appears to be essentially compatible with	the experimental results. It corresponds to the number of signal events which is below the $95\%$ C.L. upper bound divided by 1.5. Benchmark points with $0.67<r<1.5$ actually present possibly large uncertainties (originating from e.g.\ parton distribution fuction sets, the choice of renormalization and factorisation scale, the details of parton showering or the finite MC statistics) so that a conservative approach cannot classify them as excluded at $95\%$ C.L.: we will regard them as `potentially constrained'.
	
	\section{Collider searches -- Results}
	In this section, we investigate the constraints from Run-1 searches as well as preliminary results of several Run-2 searches (for an integrated luminosity of $36~\text{fb}^{-1}$) on our benchmark samples for all five scenarios presented in Subsect.~\ref{subsec:scenarios}. 
	For each model point, we generated MC events, estimated the detector response and explicitly probed the impact of the ATLAS and CMS searches summarized in Table.~\ref{tab:lhc_searches}. 
	Of course, we naively expect the constraints from Run-2 to have a higher reach than those of Run-1, due to the larger cross sections accessed at a center-of-mass energy of $13$~TeV (roughly a factor $2$ at the partonic level). On the other hand, the SM background processes also become more prominent at $13$~TeV.
	
	\subsection{MSSM-like spectra}
\begin{figure} \centering
		\includegraphics[width=0.63\textwidth,scale=0.8]{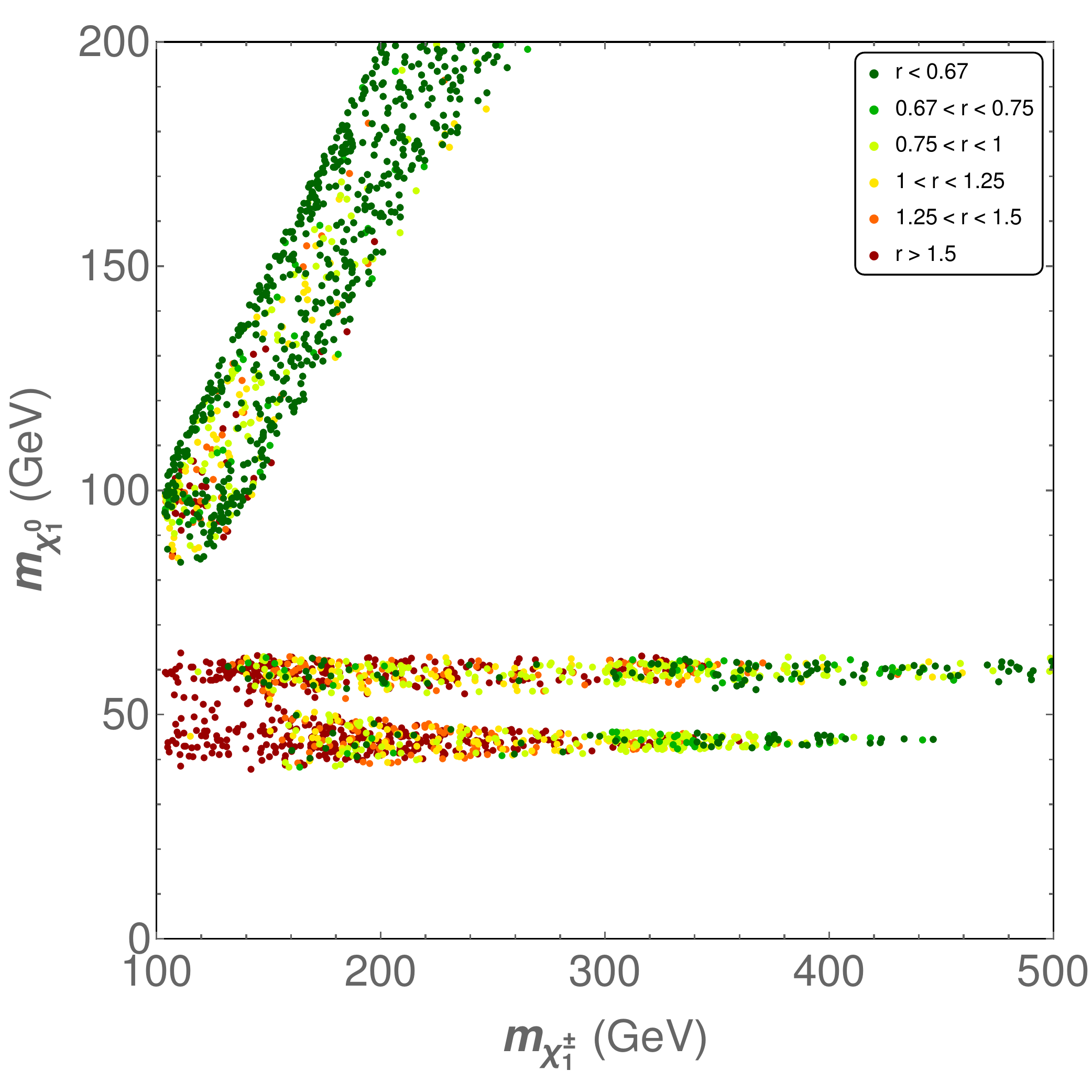} 
		\includegraphics[width=0.63\textwidth,scale=0.8]{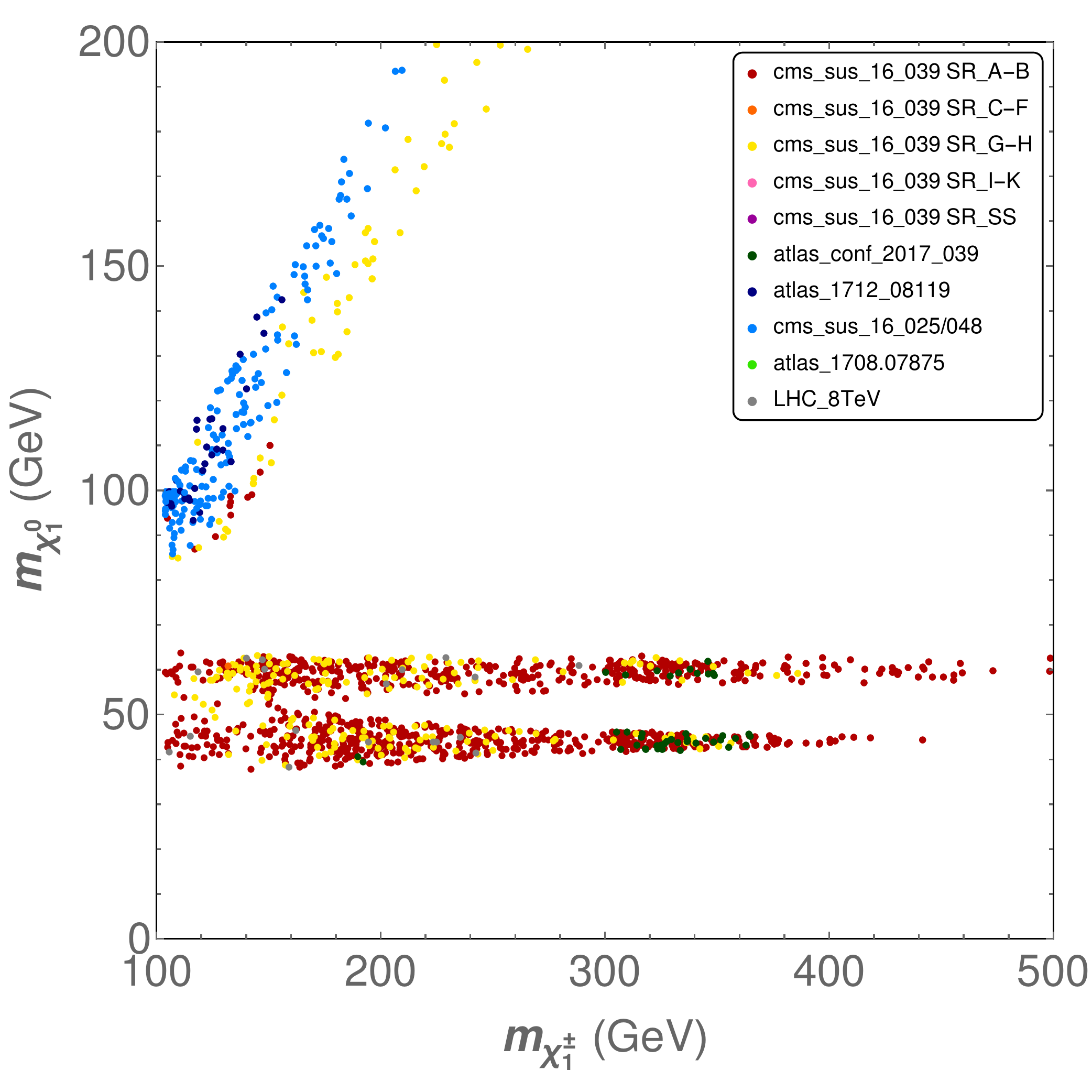}  
		\caption{Model points with MSSM-like spectra in the plane defined by the masses of $\chi_1^\pm$ and $\chi_1^0$.\\ {\sc Top:} Exclusion plot. The color of the points reflects the performance
			in view of collider searches: dark green $\to\ r<0.67$ (allowed); middle green $\to\ 0.67<r<0.75$; light green $\to\ 0.75<r<1$; yellow $\to\ 1<r<1.25$; orange $\to\ 1.25<r<1.5$; 
			red $\to\ r>1.5$ (excluded).\\ {\sc Bottom:} Constraining analyses. Each color corresponds to the category of searches that is most sensitive to the spectrum, for all points with $r>0.67$.}
		\label{fig:mssm}
	\end{figure}

	\paragraph{General discussion --}
	In Fig.~\ref{fig:mssm}, we show our randomly sampled model points in the plane defined by the masses of the lightest neutralino and chargino states: $m_{\chi_1^0}$ and 
	$m_{\chi_1^\pm}$. Chargino masses below $\sim100$~GeV are inaccessible due to LEP2 constraints. Moreover, the cross sections for chargino / neutralino production in $p-p$ collisions are suppressed in the high mass-range. Constraints in lepton searches correspondingly weaken, although the efficiency tendencially increases for large mass-splittings between the LSP and the NLSP, due to the emission of leptons with large transverse momentum.
	All the model points that we keep in this sample satisfy the constraints imposed in Subsect.~\ref{subsec:pheno_constraints} and distribute in the $\chi_1^0$/$\chi_1^{\pm}$ mass-plane according to the characteristic structures that result from the relic-density condition. The $Z$-\ and Higgs-funnel regions can be easily identified and are characterized by $m_{\chi_1^0}\approx45$~GeV and $m_{\chi_1^0}\approx60$~GeV, respectively: in these cases, DM annihilation proceeds mainly through the mediation in the $s$-channel of a $Z$ or SM Higgs bosons close to their mass-shell. In the upper left-hand corner of Fig.~\ref{fig:mssm}, the proximity in mass of the chargino and the neutralino opens the path to chargino coannihilation. 
	
	In the upper part of Fig.~\ref{fig:mssm}, we display the constraints of the collider test. Model points which are clearly allowed ($r\le0.67$) or clearly excluded ($r\ge1.5$) by direct LHC searches are shown in green or red, respectively. Ambiguous points ($0.67<r<1.5$) are depicted in intermediate shades, depending on the value of $r$. In the lower part of Fig.~\ref{fig:mssm}, we show which category of searches are most sensitive to the spectra, for points with $r>0.67$. 
	
	We observe that the collider searches constrain benchmark points in both funnel regions up to chargino masses of about $500$~GeV. However, strict exclusion ($r>1.5$) only applies for $m_{\chi_1^{\pm}}\lsim300$~GeV and even then, many points continue to satisfy $r<1.5$ and even $r<0.67$. For light charginos with mass $\lsim200$~GeV, production cross sections of charginos and neutralinos may reach $\mathcal{O}(5~\text{pb})$ for a center-of-mass energy of $8$~TeV (resp.\ $\mathcal{O}(10~\text{pb})$ at $13$~GeV), mediating sizable cross sections with $2$ or $3$ leptons in the final state. On the other hand, the reduced available phase-space tends to produce somewhat soft leptons, which reduces the efficiency of the searches. The competition of these two effects explains the inhomogeneity of the pattern of constraints in the low-mass region. However, we checked that the excluded points are usually associated with the largest leptonic cross sections for a given kinematic range (which is almost fully determined by the chargino mass, if the LSP is in the funnel region). In this region, the most efficient of the considered searches appear to be the CMS multilepton $13$~TeV searches \cite{CMS-PAS-SUS-16-039} of SR~A type (three light leptons in the final state and at least one OSSF pair). The $8$~TeV searches are largely superseded and only proved efficient for $m_{\chi_1^{\pm}}\lsim140$~GeV anyway. We find that the ATLAS $13$~TeV search tends to be less competitive than the CMS analysis. The extensive binning of the CMS search in multiple signal regions explains this situation, since it provides sensitivity to a wide range of mass scales and mass hierarchies. 
	
	In the chargino-coannihilation region (top left-hand corner), the sparticle spectrum is compressed ($m_{\chi_1^\pm}-m_{\chi_1^0}\approx 0$), so that the leptons in the final state are endowed with very little transverse momentum. The detection of these soft leptons is experimentally problematic. The effort to investigate compressed spectra at $13$~TeV results in occasional limits in our sample of points. However, as already discussed in section~\ref{sec:num_signatures}, we did not generate matched events in the compressed region, so that the results should be understood at a purely qualitative
	level. 

	As expected, the searches for three light leptons in the final state provide the best sensitivity to the spectra. The corresponding signal regions are optimized for the process $pp\rightarrow \chi_2^0\chi_1^\pm$ with subsequent decays of $\chi_2^0\rightarrow Z^{(*)}\chi_1^0$ and $\chi_1^\pm\rightarrow W^{(*)}\chi_1^0$ (the superscript $^{(*)}$ indifferently marks on-\ or off-shell gauge bosons). While the cross sections involving $\tau$ final states are competitive, we observe no sensitivity of the associated signal regions, which can be easily understood from the reduced efficiency for the identification of hadronic $\tau$'s in the detector. 

	We also separately considered the impact of $8$~TeV searches in order to compare our results with the ATLAS analysis of Ref.~\cite{Aaboud:2016wna}. The general shape of the region of exclusion are consistent, but the ATLAS study seemed to hint at somewhat more efficient collider limits in both funnel regions than those observed in our sample. These differences could be the consequence of the likelihood-driven scan performed in \cite{Aaboud:2016wna}, while we restrict ourselves to a simple random scan: thus, the sampling may depend on the priors. In addition, an extrapolation technique was applied to evaluate the limits in ref.~\cite{Aaboud:2016wna}, whereas we chose to display only the benchmark points for which we performed an actual test. Finally, \texttt{CheckMATE 2} cannot account for the correlations among signal regions. It only applies Eq.~(\ref{r}) to the most promising channel, in order to check whether a benchmark point is excluded or not. This method is expected to be more conservative than the ATLAS procedure, which consists in calculating the $p$-value. 
	
	\paragraph{Test-points --}
	In Table~\ref{MSSMexample}, we list a few points illustrating various features of the MSSM-like scenario.
	{\begin{table}[t]
			\begin{tabular}{|c||c|c|c|c|c}
				\hline
				& \verb|39_A18|                  & \verb|52_A5|                   & \verb|spectr31|                & \verb|139_A24|                 &  \\ \hline\hline
				$m_{\chi_1^0}$ (GeV)     & B\ \ \ \ \ $43$                & B\ \ \ \ $50$                  & B\ \ \ \ \ $43$                & B\ \ \ \ \ $41$                &  \\ \hline
				$m_{\chi_2^0}$ (GeV)     & W\ \ \ \ $125$                 & W/H\ $116$                     & W\ \ \ \ $222$                 & H\ \ \ \ $192$                 &  \\ \hline
				$m_{\chi_1^{\pm}}$ (GeV) & W\ \ \ \ $125$                 & W/H\ $117$                     & W\ \ \ \ $223$                 & H\ \ \ \ $190$                 &  \\ \hline\hline
				BR$[\chi_1^{\pm}\to\chi^0_1W]$ & $1$                        & $1^*$                          & $1$                            & $1$                            &  \\ \hline
				BR$[\chi_i^0\to\chi_1^0Z]$ & $1^*$ (i=2)                  & $1^*$ ($i=2$)                  & $0.97$ ($i=2$)                 & $0.58_{i=2}$; $0.85_{i=3}$     &  \\ \hline\hline
				$\sigma_{\text{8\,TeV}}[pp\to\chi_i^0\chi_1^{\pm}]$ (pb) & $2.45$\ \ ($i=2$) & $2.02$\ \ ($i=2$)           & $0.27$\ \ ($i=2$)              & $0.31$\ \ ($i=2+3$)            &  \\ \hline
				$\sigma_{\text{8\,TeV}}[pp\to\chi_1^+\chi_1^-]$ (pb) & $1.27$                & $1.13$                      & $0.13$                         & $0.10$                         &  \\ \hline
				$\sigma_{\text{13\,TeV}}[pp\to\chi_i^0\chi_1^{\pm}]$ (pb) & $4.94$ (i=2)   & $4.02$ (i=2)                   & $0.62$ (i=2)                   & $0.69$ (i=2+3)                 &  \\ \hline
				$\sigma_{\text{13\,TeV}}[pp\to\chi_1^+\chi_1^-]$ (pb) & $2.59$             & $2.28$                         & $0.31$                         & $0.22$                         &  \\ \hline\hline
				$\sigma_{\text{8\,TeV}}[pp\to3\ell]$ (fb)   & $45$                     & $43$                      & $5$                      & $4$                           &  \\ \hline
				$\sigma_{\text{13\,TeV}}[pp\to3\ell]$ (fb)   & $93$                          & $89$                          & $12$                           & $9$                           &  \\ \hline\hline
				Search                   & { SR A$08$}       & { SR A$07$}       & { SR A$29$}       & { SR A$25$}       &  \\ \hline
				$r$                      & $4.7$                          & $2.1$                          & $2.5$                          & $0.5$                          &  \\ \hline
			\end{tabular}\\
			\null\hfill\begin{tabular}{c|c|c|c|}
				\hline
				& \verb|295_A24|                 &  \verb|11_A39|                 & \verb|10_A41A|                 \\ \hline\hline
				& B\ \ \ \ \ $59$                & B/H\ \ $95$                    & H\ \ \ \ \ $99$                \\ \hline
				& H\ \ \ \ $152$                 & H\ \ \ \ $123$                 & H\ \ \ \ $112$                 \\ \hline
				& H\ \ \ \ $150$                 & H\ \ \ \ $113$                 & H\ \ \ \ $104$                 \\ \hline\hline
				& $1$                            & $1^*$                          & $1^*$                          \\ \hline
				& $1$ ($i=2,3$)                  & $1^*$ ($i=2,3$)                & $0.89^*$ ($i=2$)            \\ \hline\hline
				& $0.69$\ \ ($i=2+3$)            & $1.06$\ \ ($i=2+3$)            & $1.39$\ \ ($i=2$)              \\ \hline
				& $0.23$                         & $0.66$                         & $0.89$                         \\ \hline
				& $1.45$ (i=2+3)                 & $2.13$ (i=2+3)                 & $2.72$ (i=2)                   \\ \hline
				& $0.48$                         & $1.31$                         & $1.73$                         \\ \hline\hline
				& $11$                           & $18$                      & $24$                            \\ \hline
				& $23$                           & $35$                           & $47$                           \\ \hline\hline
				& { SR A$25$}       & { SR1-wk-1l-mll2} & { SR1-wk-1l-mll1} \\ \hline
				& $0.6$                          & $0.5$                          & $3.2$                          \\ \hline
			\end{tabular}
			\caption{Test-points in the MSSM-like scenario. The first lines detail the characteristics of the lightest electroweakino states, with B, W, H and S standing for bino, wino, higgsino and singlino, and documenting the nature of the dominant component of the considered state. In case of a large mixing, both large components are indicated. For the other scenarios, we will also provide the masses of the lightest Higgs states, but in the case of the MSSM-like scenario, only the SM-like Higgs is light, with mass 				$\sim125$~GeV, so that we skip such information here. We then provide the magnitude of the `traditionally' leading branching fractions into $W$, $Z$ for the light charginos and neutralinos. There, the superscript $^*$ indicates that the decay is mediated off-shell. Any departure from $1$ implies the existence of other important decay channels: in this particular scenario, these correspond essentially to decays involving the SM-like Higgs, but in the other scenarios, decays involving the Higgs singlets, the singlino or even the photon can be relevant. Then appear the cross sections of the leading production channels in $p-p$ collisions for $8$~TeV as well as $13$~TeV center-of-mass energy. Next, we derive the value of the cross sections mediated by the electroweakinos for $3$ light leptons in the final state. We do not indicate the cross sections involving $\tau$ final states since we observe no (or little) experimental sensitivity to them. Finally, the last two lines identify the name of the most relevant search in the analysis of \texttt{CheckMATE} and the associated value of the ratio $r$: below $0.67$, we regard the point as allowed, and excluded above $1.5$; intermediary values mean that the signal is close to exclusion but that a full accounting of uncertainties would certainly place it within error bars. \label{MSSMexample}}
			\vspace{-6ex}
	\end{table}}\afterpage{\clearpage}
	\begin{itemize}
		\item \verb|39_A18| involves a light electroweakino sector, with a bino LSP in the $Z$-funnel, winos at a mass of $\sim125$~GeV and higgsinos at $\sim250$~GeV. The corresponding production cross sections at the LHC are rather large and reach up to $\sim2$~pb (resp.\ $5$~pb) in the $\chi_1^{\pm}\chi_2^0$ or $\chi^+_1\chi^-_1$ channels at a center-of-mass energy of $8$~TeV (resp.\ $13$~TeV). The wino decays are then essentially mediated by electroweak gauge bosons, leading to $BR[\chi^+_1\to\chi^0_1 \ell^+\nu_\ell]\simeq22\%$ and $BR[\chi^0_2\to\chi^0_1 \ell^+\ell^-]\simeq6.6\%$, where $\ell$ denotes light leptons. These channels turn out to be the main contributors to the $pp\to3\ell$ cross sections and result into predicted signals of $\sigma_{\text{8\,TeV}}[pp\to3\ell]\simeq45$~fb and $\sigma_{\text{13\,TeV}}[pp\to3\ell]\simeq93$~fb. As a result, $3\ell$ searches are quite efficient. In fact, the corresponding ATLAS Run-1 searches of type SR$\tau$a are sufficient to conclude to the exclusion of this point. These limits are superseded by the Run-2 results, where the CMS trilepton search is the most powerful one. The most sensitive signal region, SRA08, requires one OSSF dilepton pair and moderate cuts on transverse missing momentum ($150\le\met\le 200$ GeV), transverse mass ($100\le m_T\le160$ GeV) and invariant dilepton mass ($m_{\ell\ell}\le 70$ GeV). For this specific point, the targeted invariant mass, corresponding to $\sim m_{\chi_2^0}-m_{\chi_1^0}$, is slightly below the $Z$-mass window, and the $\chi_2^0$ decay proceeds through an off-shell $Z$-boson. The $WZ$ signal regions of the ATLAS 13 TeV multilepton study does not show good sensitivity in this context, due to rather strict cuts and a coarse binning. On the other hand, $m_{\chi_1^{\pm}}-m_{\chi_1^0}\simeq80~\text{GeV}\simeq m_{\chi_2^0}-m_{\chi_1^0}$. This difference is close to the $W$ mass and thus the $m_{T2}$ variable is not efficient enough to suppress the $WW$ SM background for dilepton final states. As a result, the efficiency in $2\ell$ searches is rather weak. This point, rather typical of the targets of the $3\ell$-searches, is thus clearly excluded.
		\item \verb|52_A5| shares some similarities with the previous benchmark point: the bino LSP is in the funnel region and the wino and higgsino states are relatively light, with masses of order $\sim115-220$~GeV. Contrary to \verb|39_A18|, however, the spectrum is more compressed and involves a sizable wino-higgsino mixing. The latter somewhat reduces the production cross sections compared to the case of pure wino. In addition, $m_{\chi_1^{\pm}}-m_{\chi_1^0}\simeq65~\text{GeV}\simeq m_{\chi_2^0}-m_{\chi_1^0}$ and the decays occur via off shell SM gauge bosons, which means that the leptons in the final state are relatively soft. Consequently, the efficiency in searches is reduced. $8$~TeV searches are only mildly sensitive to this point (which is also related to an upward fluctuation in some signal regions targeting small mass differences $m_{\chi_2^0}-m_{\chi_1^0}$) and only Run-2 searches are actually able to exclude it. Again, the CMS multilepton search is the most efficient among the considered channels, with SRA07 being the most sensitive signal region. The cuts are very similar to SRA08 with somewhat weaker conditions ($100\le\met\le150$ GeV). This evolution with respect to the previous spectrum is expected, since the considered masses are smaller and the whole spectrum is somewhat more compressed.
		\item \verb|spectr31|: For this point, the mass of the dominantly-bino LSP falls in the Z-funnel. The wino states share a mass of $\sim220$~GeV while the higgsinos are heavier at $\sim335$~GeV. Due to the larger masses involved in the electroweakino sector, the production cross sections in $p-p$ collisions are modest as compared to the previous points (a few $100$~fb for the dominant $\chi_2^0\chi_1^+$ and $\chi_1^+\chi_1^-$ channels). Correspondingly, the lepton + MET cross sections are also reduced ($\mathcal{O}(10)$~fb). However, the produced leptons are energetic and lead to a good efficiency in searches. Constraints from Run-2 return a clear exclusion. The most sensitive signal region is SRA29. It requires a similar cut on the missing transverse momentum as SRA07 and SRA08, but the transverse cut of $m_T\ge160$ GeV is harder and the invariant mass of the OSSF dilepton pair of $75\le m_{\ell\ell}\le105$ GeV corresponds to the signal region with on shell $Z$. Most of the SM background events are expected in this $Z$-mass window. However, a hard cut on $m_T$ is fairly  efficient since roughly $86\%$ of the SM $WZ$ events of the $Z$-mass window cluster in the region defined by $m_T\le100$ GeV and $35$ GeV $\le \met\le100$ GeV are discarded. Surprisingly, even for this benchmark point, the ATLAS trilepton search is not performing very well. However, the ATLAS 
		dilepton signal regions show relatively good sensitivity to this spectrum.
		\item \verb|139_A24| also contains a bino LSP in the $Z$-funnel. The electroweakino sector includes higgsino states at $\sim190$~GeV and wino states at $\sim900$~GeV. Correspondingly, the production cross sections of charginos and neutralinos is somewhat reduced (as compared to wino states). The decay chains are dominated by $\chi_1^{\pm}\to\chi_1^0W^{\pm}$, $\chi_2^0\to\chi_1^0Z$ and $\chi_2^0\to\chi_1^0H_{\text{SM}}$, the latter channel depleting the final state of light leptons.	In the end, the \texttt{CheckMATE} analysis concludes to no sensitivity to this spectrum from both the Run-1 and Run-2 searches. In the case of the $13$~TeV searches, we should mention that the signal region that is selected as the most sensitive one, SRA25, shows an upward fluctuation in the number of observed events. As a result, the observed limit on the number of signal events is weaker than expected.
		\item \verb|295_A24|: For this point, the bino LSP is in the SM-Higgs funnel, with higgsino states around $\sim150$~GeV and winos at $\sim1$~TeV. The higgsino production in $p-p$ collisions is comparatively reduced. In addition, the mass difference $m_{\chi_2}-m_{\chi_1}$ roughly corresponds to the $Z$-mass. In this region of parameter space, events with pair production of chargino and neutralino look very similar to the dominant $WZ$ SM background. This benchmark point is therefore very challenging experimentally.
		\item \verb|11_A39| is representative of the chargino coannihilation region, with a higgsino-bino admixture as the LSP, at a mass of $\sim95$~GeV, and higgsino states in the $110-120$~GeV region. The winos are much heavier, with masses around $730$~GeV. The dominant production channels are $\chi_1^{\pm}\chi_2^0$, $\chi^+_1\chi^-_1$ and $\chi_1^0\chi_2^0$ and generate soft leptons in the final state. This results into very low efficiencies for leptonic searches. According to \texttt{CheckMATE}, this point is not constrained by multi-lepton searches. Unsurprisingly. the most sensitive channel corresponds to the CMS searches for compressed spectra \cite{CMS:2017fij,Sirunyan:2018iwl}. Comparing with the CMS exclusion plot, the benchmark points seems to be on the exclusion boundary. However, the CMS plot assumes wino pair production with subsequent decays via off shell $W$ and $Z$ boson, while the point under consideration has a 
		higgsino NLSP with reduced production cross section. We stress that we did not match the MC events, so that the theoretical uncertainty on this benchmark point could be relatively large. In particular, the CMS soft dilepton study demands the presence of one jet although the cut is very mild, with $p_T(j_1)\ge25$ GeV.
		\item \verb|10_A41A| is another example in the coannihilation region. The LSP and NLSP's are higgsino states. Bino and winos are much heavier (beyond $500$~GeV). In this case, the searches for a soft lepton pair in the final state that have been performed in Run-2 appear to be sensitive to the production and decays of the higgsinos. \texttt{CheckMATE} identifies a CMS search as the most constraining limit and concludes to the exclusion of this spectrum. The main difference with the previous benchmark point rests with the smaller mass-splitting between the LSP and the NLSP. With decreasing mass difference, the cut $0.6\le\met/H_T\le1.4$ becomes more efficient. Here, $H_T$ is the scalar sum of all jets (mainly from initial state radiation). A smaller mass gap leads to less energetic decay products of the NLSP, which tend to increase the overall missing transverse momentum.
	\end{itemize}

	\subsection{Singlino LSP}
	Now we study how the LHC electroweakino searches perform in scenarios beyond the MSSM. Thus, we first focus on points with singlino LSPs. 
	
	\begin{figure} \centering
		\includegraphics[width=0.67\textwidth,scale=0.8]{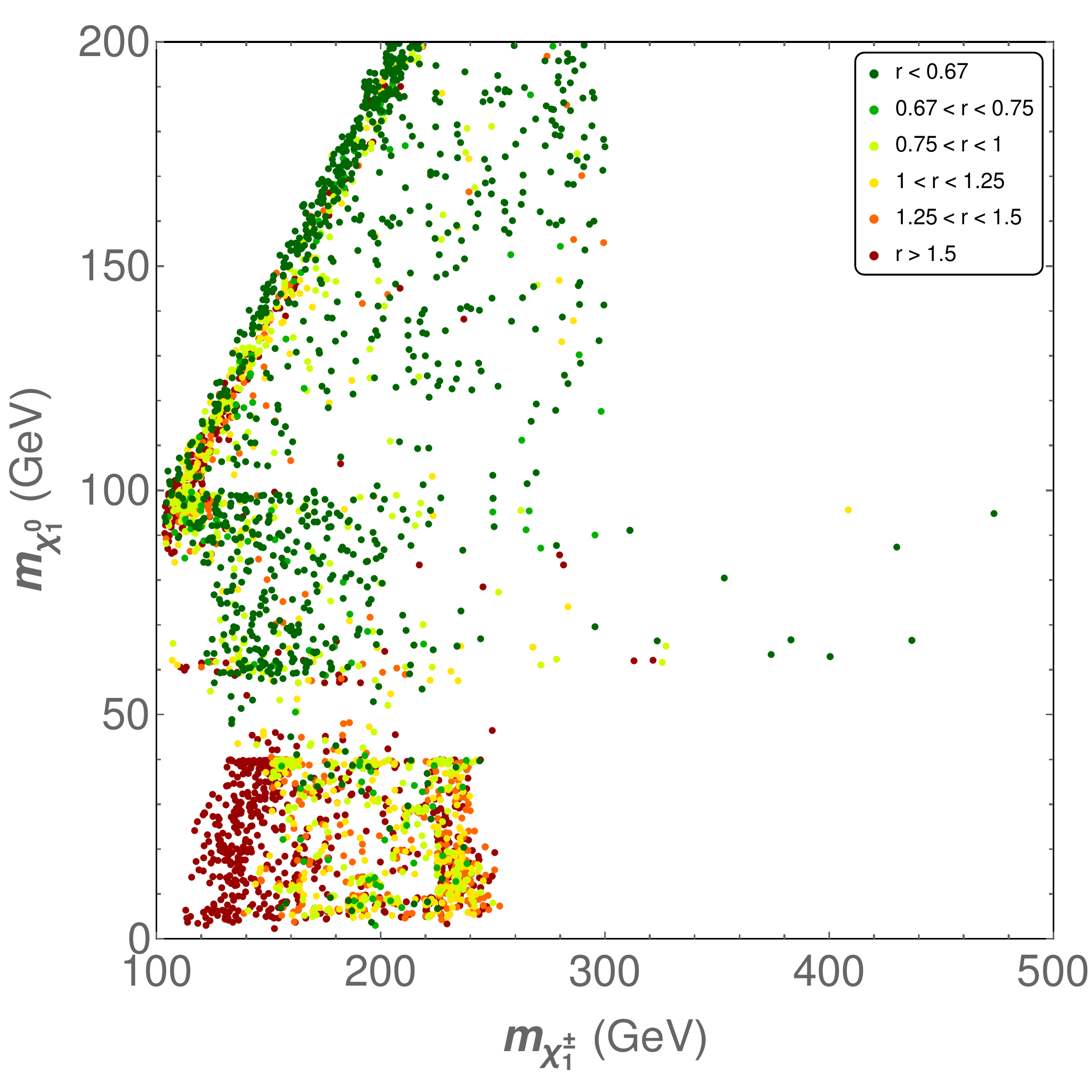} 
		\includegraphics[width=0.67\textwidth,scale=0.8]{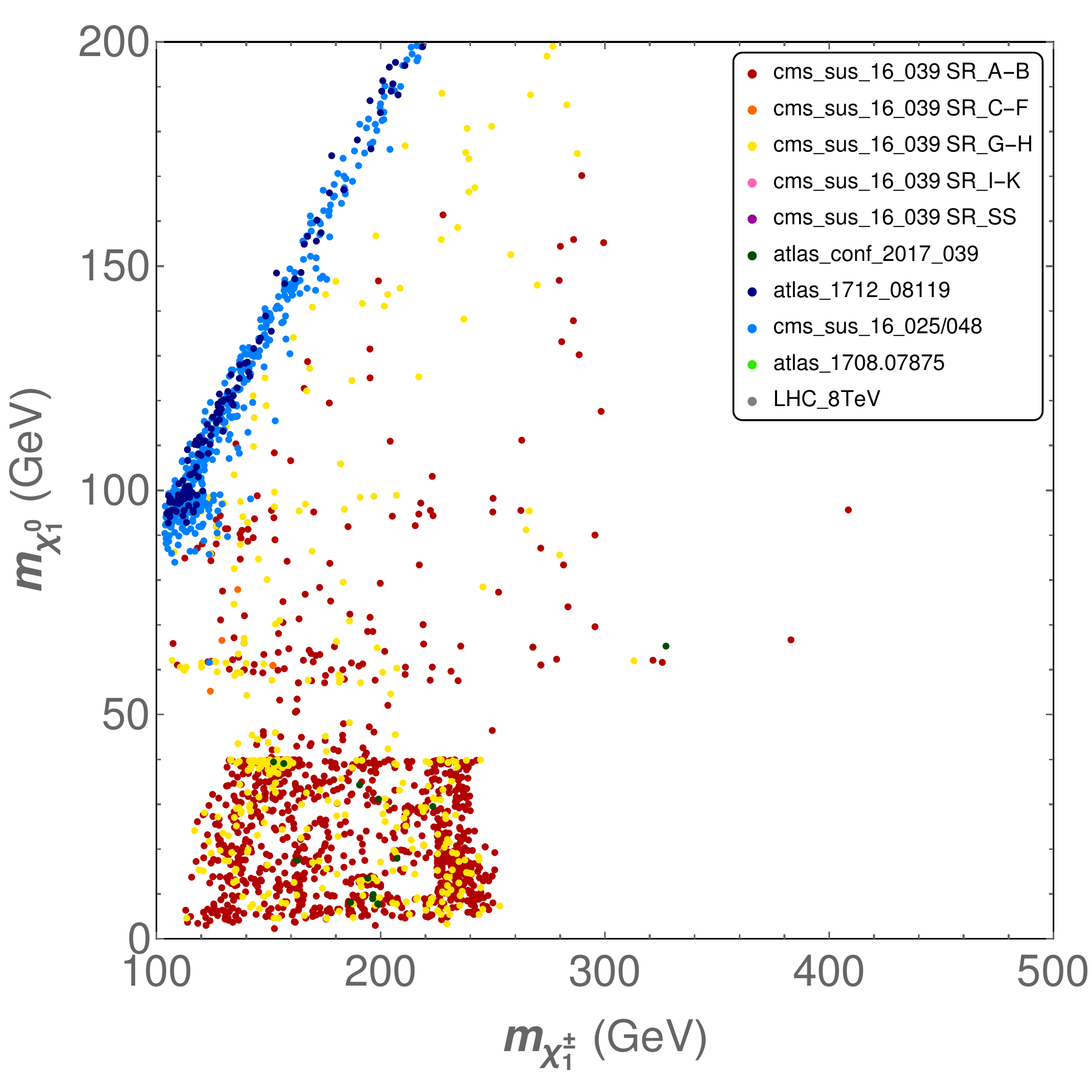} 
		\caption{Impact of the collider searches on the benchmark points involving a singlino-like LSP. The results are shown in the plane defined by the masses of the 
			lightest neutralino and chargino states. The color code is the same as in Fig.~\ref{fig:mssm}.}
		\label{fig:singlino_lsp}
	\end{figure}

	\paragraph{Global analysis --}
	Our results for the collider searches are shown in Fig.~\ref{fig:singlino_lsp}. The color labels are the same as in Fig.~\ref{fig:mssm}. In contrast with the previous subsection, the model points distribute across most of the mass plane, with LSP masses as low as $2$~GeV. The possible existence of light singlet scalar states funneling the DM annihilation in the early Universe explains the traditional `funnel' and `coannihilation' structures fading away. Contrarily to \cite{Mou:2017sjf}, we find spectra involving LSP masses below $\sim10$~GeV and satisfying limits from the Higgs and $Z$ decays. This scenario is often considered as a trademark of the NMSSM. On the other hand, we find few benchmark points of this type with both a light LSP and a heavier NLSP 	($m_{\chi^{\pm}_1}\gsim300$~GeV). This is in fact a relatively generic feature of the $Z_3$-conserving NMSSM: in this model, it is indeed difficult to simultaneously satisfy stability conditions (positive Higgs squared masses) together with the relic density constraint (implying $m_{\Phi}\sim2m_{\chi_1^0}$, where $m_{\Phi}$ represents the mass of a mostly singlet CP-even or CP-odd Higgs state) when requiring a large value of $\mu_{\text{eff}}$ (for a heavy $\chi_1^{\pm}$).
	
	In the upper plot of Fig.~\ref{fig:singlino_lsp}, most benchmark points that are excluded by the leptonic searches have LSP masses below 40 GeV. Even there, we continue to observe outlier points with $r<0.67$. The impact of the multi-lepton searches on this scenario thus appears limited, although the corner with $m_{\chi_1^0}\lsim40$~GeV and $m_{\chi_1^{\pm}}\lsim150$~GeV looks strongly constrained. As compared to the MSSM-like scenario, we actually observe that, for comparable kinematical configurations, the production cross sections of electroweakinos (hence the mediated $2\ell$ and $3\ell$ cross sections) tend to be reduced by $\sim20\%$ in the scenario with singlino-LSP. We understand this fact as a consequence of the characteristics of the spectra that are selected by the relic-density condition, due to the different nature of the LSP (singlino vs.\ bino), and the systematic presence of a (comparatively) light higgsino component. This means that somewhat larger luminosities are needed in order to achieve comparable limits as those of the MSSM-like scenario. We also note that, due to the opening of the Higgs-singlet funnels mediating DM annihilation, the kinematical configurations can be relatively different from those encountered in the MSSM-like scenario. On the other hand, the decays of the neutralinos and charginos 
	are largely unchanged (at the level of the branching ratios), although the presence of light Higgs singlet states ($\Phi$) occasionally induces decays of the form $\chi_i^0\to\Phi\chi_1^0$. According to the analysis of \texttt{CheckMATE}, CMS $13$~TeV searches for three light leptons (SR~A) or more (SR~G) are currently placing the most effective limits, except in the coannihilation region, where dedicated searches for compressed spectra perform better. On the whole, however, many points with light spectra are compatible with the collider constraints. This scenario with light singlino-like LSP thus remains phenomenologically viable from the perspective of collider searches. Previous studies \cite{Ellwanger:2016sur} have insisted on the resilience of this scenario even to $3000~\text{fb}^{-1}$ searches at a HL-LHC, although the corner involving light LSPs and heavy charginos should be covered.

	\paragraph{Test-points --}
	In Table~\ref{singlinoLSPexample}, we present various examples of points with singlino LSP.
	\begin{table}[t]
		\begin{tabular}{|c||c|c|c|c|c}
			\hline
			& \verb|11_B3A|                  & \verb|2_B5|                    & \verb|133_B5A|                 & \verb|33_B3A|                  & \\ \hline\hline
			$m_{\chi_1^0}$ (GeV)     & S\ \ \ \ \ \ $7$               & S\ \ \ \ \ \ $8$               & S\ \ \ \ \ $24$                & S\ \ \ \ \ $29$                & \\ \hline
			$m_{\chi_2^0}$ (GeV)     & H\ \ \ $167$                   & H\ \ \ $158$                   & H/W\ $131$                     & H/B\ $141$                     & \\ \hline
			$m_{\chi_1^{\pm}}$ (GeV) & H\ \ \ $171$                   & H\ \ \ $158$                   & H/W\ $130$                     & H\ \ \ $162$                   & \\ \hline\hline
			$m_{H_1^0}$ (GeV)        & $15$                           & $18$                           & $20$                           & $11$                           & \\ \hline
			$m_{A_1^0}$ (GeV)        & $24$                           & $9$                            & $49$                           & $63$                           & \\ \hline\hline
			BR$[\chi_1^{\pm}\to\chi_1^0W]$ & $1$                      & $1$                            & $1$                            & $1$                            & \\ \hline
			BR$[\chi_i^0\to\chi_1^0Z]$ & $0.86_{i=2}$; $0.54_{i=3}$   & $0.90_{i=2}$; $0.59_{i=3}$    & $0.87_{i=2}$; $0.78_{i=3}$     & $0.83_{i=2}$; $0.80_{i=3}$     & \\ \hline\hline 
			$\sigma_{\text{8\,TeV}}[pp\to\chi_i^0\chi_1^{\pm}]$ (pb) & $0.47$ ($i=2+3$) & $0.21$ ($i=2+3$)             & $1.86$ ($i=2+3$)               & $0.46$ ($i=2+3$)               & \\ \hline
			$\sigma_{\text{8\,TeV}}[pp\to\chi_1^+\chi_1^-]$ (pb) & $0.14$             & $0.19$                         & $0.72$                         & $0.17$                         & \\ \hline
			$\sigma_{\text{13\,TeV}}[pp\to\chi_i^0\chi_1^{\pm}]$ (pb) & $1.01$ ($i=2+3$) & $0.47$ ($i=2+3$)             & $2.75$ ($i=2+3$)               & $0.98$ ($i=2+3$)               & \\ \hline
			$\sigma_{\text{13\,TeV}}[pp\to\chi_1^+\chi_1^-]$ (pb) & $0.31$             & $0.41$                         & $1.48$                         & $0.38$                         & \\ \hline\hline
			$\sigma_{\text{8\,TeV}}[pp\to3\ell]$ (fb)   & $6$                      & $3$                            & $25$                      & $7$                           & \\ \hline
			$\sigma_{\text{13\,TeV}}[pp\to3\ell]$ (fb)   & $12$                           & $6$                           & $53$                          & $14$                          & \\ \hline\hline
			Search                   & { SR A$29$}       & { SR A$25$}       & { SR G$05$}       & { SR A$25$}       & \\ \hline
			$r$                      & $1.2$                          & $0.5$                          & $4.6$                          & $0.6$                          & \\ \hline
		\end{tabular}\\
		\null\hfill\begin{tabular}{c|c|c|c|c|}
			\hline
			& \verb|82_B4|                   & \verb|98_B5A|                  & \verb|118_B5|                  & \verb|46_B5|                   \\ \hline\hline
			& S\ \ \ \ \ $37$                & S\ \ \ \ \ $40$                & S/H\ \ $63$                    & S/H\ \ $64$                    \\ \hline
			& W\ \ \ $136$                   & H\ \ \ $133$                   & H\ \ \ $146$                   & H\ \ \ $146$                   \\ \hline
			& W\ \ \ $134$                   & H\ \ \ $133$                   & H\ \ \ $137$                   & H\ \ \ $122$                   \\ \hline\hline
			& $37$                           & $18$                           & $144$                          & $77$                           \\ \hline
			& $83$                           & $65$                           & $138$                          & $70$                           \\ \hline\hline
			& $1$                            & $1$                            & $1^*$                          & $1$                            \\ \hline
			& $0.97_{i=2}$                   & $0.91_{i=2}$; $0.93_{i=3}$     & $1^*_{i=2}$; $0.45_{i=3}$      & $0.07_{i=2}$; $0.06_{i=3}$     \\ \hline\hline
			& $1.70$ ($i=2$)                 & $1.12$ ($i=2+3$)               & $0.64$ ($i=2+3$)               & $0.29$ ($i=2+3$)               \\ \hline
			& $0.94$                         & $0.47$                         & $0.33$                         & $0.50$                         \\ \hline
			& $3.47$ ($i=2$)                 & $2.30$ ($i=2+3$)               & $1.34$ ($i=2+3$)               & $0.68$ ($i=2+3$)               \\ \hline
			& $1.94$                         & $0.98$                         & $0.68$                         & $0.35$                         \\ \hline\hline
			& $31$                      & $20$                      & $9$                      & $0$                            \\ \hline
			& $65$                          & $44$                          & $18$                           & $1$                           \\ \hline\hline
			& { SR A$24$}       & { SR G$03$}       & { SR A$03$}       & { SR F$04$}   \\ \hline
			& $2.8$                          & $3.7$                          & $0.8$                          & $0.4$                          \\ \hline
		\end{tabular}
		\caption{Test-points for the singlino LSP scenario. The general features are similar to those of Table~\ref{MSSMexample}. \label{singlinoLSPexample}}
		\vspace{-2ex}
	\end{table}
	\begin{itemize}
		\item \verb|11_B3A| represents a first example of a very-light dark-matter candidate, with mass $\sim7$~GeV (singlino at $\sim98\%$). An efficient annihilation of this particle is only possible because of the existence of a light Higgs mediator $H_1$ with mass of $\sim15$~GeV. The higgsinos, bino and winos take masses of about $170$, $350$ and $1000$~GeV respectively. The production cross sections of these states at the LHC are not particularly large (suppression by the mass-scale). On the other hand, the mass-splitting ensures the presence of energetic leptons in the final state. Although both a light CP-odd and a light CP-even Higgs states are present, the decays of the higgsinos and winos essentially involve the EW gauge and the SM-like Higgs bosons. According to \texttt{CheckMATE}, the CMS search for three light leptons at Run-2 returns $r=1.2$. The signal region with best sensitivity is SRA29. The large mass-splitting between the LSP and the NLSP explains the selection of a signal region with a rather strict cut on the transverse mass $m_T\ge160$ GeV, while the cut on missing transverse momentum $200\le\met\le250$ GeV is rather moderate. Further statistics should place this point within the range of clear exclusion. 
		\item \verb|2_B5| is similar to the previous point, with a singlino LSP at $\sim8$~GeV, higgsino states at $\sim160$~GeV, a bino at $\sim500$~GeV and the winos at $\sim1$~TeV. The higgsino production cross section is smaller than that of the previous point, resulting in smaller $3\ell$ signals. Both a light CP-even and a light CP-odd singlets are present, but they only have a moderate impact on higgsino decays. In addition, due to the somewhat reduced mass-splitting between the lightest neutralino and the lighter chargino, signal events are unable to satisfy a very tight cut on the transverse mass, so that the most sensitive signal region is SRA25, with a somewhat weaker cut on $m_T$ as compared to the selected signal region of the previous benchmark point. The \texttt{CheckMATE} analysis concludes to the absence of constraints from multilepton searches.
		\item \verb|133_B5A| includes a singlino-like LSP with mass $\sim24$~GeV, annihilating in the $A_1$-funnel, higgsino and wino (mixed) states at $\sim130$ and $\sim265$~GeV, and a heavy bino. The production cross section of electroweakinos is of order $\sim1$~pb in the dominant channels, leading to sizable $3\ell$ cross sections. However, these are tendentially reduced as compared to those of comparable spectra in the MSSM-like scenario. On the other hand, the mass-splitting between the singlino and the NLSP is also larger compared to {\it similar} MSSM scenarios, since the singlino LSP is lighter than the typical bino in the $Z$-funnel. As a result, the leptons are more energetic and the reach of the light lepton searches is higher. According to \texttt{CheckMATE}, already the Run-1 $3$-lepton search {SR0$\tau$a$16$} is sensitive to this point, with $\sim10$ predicted events against an experimental limit of $\sim6$, hence leading to tensions. Full exclusion is achieved at Run-2. Interestingly, the most sensitive signal region is a CMS search for more than $3$ light leptons in the final state, labeled with G. Indeed, the higgsino spectrum allows for $\chi^0_2\chi^0_3$ production with subsequent decays into two $Z$ bosons. Despite the small branching ratio for a 4 light-lepton final state, the signal is almost background-free and thus leads to a viable search. The signal region G05 explicitly demands 4 light leptons with the further requirement of two OSSF pairs and a tight cut on the missing transverse momentum $\met\ge200$ GeV, with the goal to further suppress the SM background processes. 
		\item \verb|33_B3A| contains a singlino LSP at $\sim29$~GeV, in the $A_1$ annihilation funnel. The NLSP is a bino/higgsino admixture at $\sim140$~GeV while the other higgsino states are somewhat heavier ($\sim160$~GeV) and the winos are at $\sim1$~TeV. The production cross sections of the bino/higgsino states as well as the $3\ell$ cross sections are comparable to those of \verb|11_B3A|. However, the decays of the  bino/higgsinos involve the light Higgs singlets at the level of $\sim15\%$ and the mass-splitting between LSP and NLSP is reduced. \texttt{CheckMATE} finds no conclusive sensitivity for both Run-1 and Run-2 searches.
		\item \verb|82_B4|: This point has loose similarities with \verb|39_A18| of the MSSM-like subset. The mass of the singlino LSP falls in the $A_1$-funnel but is also quite close to the $Z$-funnel. The NLSP's are wino-like, with masses of $\sim135$~GeV, while the higgsinos take masses in the range $\sim250-290$~GeV.	The decays of $\chi_{2,3,4}^0$, $\chi_1^{\pm}$ are essentially mediated by gauge bosons. However, the production cross sections ($\chi_1^+\chi_1^-$, $\chi_1^+\chi_2^0$) at the LHC and the subsequent multi-lepton signals are reduced by $\sim30\%$ as compared to \verb|39_A18|. Only Run-2 results are thus able to exclude this point, yet with a lower significance as for \verb|39_A18|. 
		\item \verb|98_B5A|: This point differs from the previous one in that the light NLSP's with mass around $\sim133$~GeV are higgsino-like, while the winos take mass in the range of $300$~GeV. The production cross section of the higgsinos in proton-proton collisions is further reduced. As before, light singlet Higgs states are present in the spectrum and allow for an efficient annihilation cross section of the singlino-like NLSP, irrespectively of the very low electroweak charge of this state (gauge-singlet at $\sim95\%$). On the other hand, the  decays of the SM-like Higgs or the electroweakinos into singlet Higgs states remain at the percent level, hence have little impact on the chargino/neutralino phenomenology at colliders. The decays of the electroweakinos is still dominated by the mediation of $W$ and $Z$. Exclusion is achieved at Run-2, via the CMS search for more than $4$ light leptons in the final state. This benchmark has some similarity with \verb|133_B5A|: in both cases, higgsinos have the same mass-scale but the mass difference is smaller for \verb|98_B5A|, thus barely allowing for an on-shell decay of both $Z$ bosons in the decays of a produced higgsino pair. As a consequence, less net missing transverse momentum is expected and the best sensitive signal region G03 has a weaker requirement, with $100\le\met\le150$ GeV.  
		\item \verb|118_B5|: For this point, the LSP is in the SM-Higgs funnel. The singlet Higgs states do not play a critical role for the annihilation cross section. Although dominantly singlino at $\sim56\%$, the lightest neutralino contains a sizable higgsino component. Further higgsino states come with a mass of $\sim140$~GeV while the winos are heavy ($\sim1$~TeV). This point is only mildly constrained by the Run-2 $3\ell$ searches. The small production cross section of higgsinos is combined with a small mass-splitting, which results in decays through off-shell $Z$ and $W$ bosons, hence in a poor sensitivity of all Run-2 searches. Interestingly,  $2\ell$ searches at Run-1 ({WWc SF}) place stronger bounds, which would result in $r=1.25$. However, this is the consequence of a downward fluctuation in the corresponding signal region: the expected $S_{95}$ limit is actually smaller than the observed one. According to the selection rules of the most sensitive search by \texttt{CheckMATE}, the CMS Run-2 $3\ell$ search A03 is chosen, instead, as the relevant test-channel, leading to the weaker limit $r=0.8$.
		\item \verb|46_B5| also includes a singlino/higgsino LSP in the SM-Higgs funnel, with the remaining higgsinos in the range $120-150$~GeV and the binos and winos around $0.5$~TeV. The cross sections are somewhat weaker than for the previous point and the higgsino decays dominantly involve the singlet-Higgs states with mass $\sim70-77$~GeV. In turn, these Higgs states dominantly decay into bottom pairs. The lepton flavor appearing in Higgs decays is essentially $\tau$, reducing the multiplicity of light leptons in the final state. According to \texttt{CheckMATE}, no tension with the Run-1 or Run-2 searches exists for this point. As no SUSY search is optimized for decay chains ending in hadronically decaying light scalar states and that $\tau$-searches are less efficient than those involving light leptons in the final state, this point is difficult to probe. The best `sensitivity' is obtained for the CMS $13$~TeV search with a $\tau$ pair and a light lepton.
	\end{itemize}
	
	\subsection{Singlino NLSP}
	In this subsection, we consider	the impact of the LHC searches on a scenario with light singlino NLSP. 
	
	\paragraph{Global analysis --} In this scenario, the lightest neutralino typically is almost systematically bino-dominated. In the plane $(m_{\chi^0_1},m_{\chi^{\pm}_1})$, the $Z$- and SM Higgs-funnel structures again emerge as a consequence of the condition on the thermal DM relic abundance. The low-mass range for the chargino ($100-150$~GeV) proves to be less populated in this sample than in the MSSM case: this can be understood as a consequence of the requirement for an intermediary singlino-like NLSP. The coannihilation region is largely irrelevant in this scenario, since the spectrum would imply an intermediary singlino state between the already almost degenerate wino or higgsino neutralino and chargino. However, we also find occasional points where the upper bound for the thermal DM relic density can be satisfied through the mediation of a scalar singlet.
	
	\begin{figure} \centering
		\includegraphics[width=0.67\textwidth,scale=0.8]{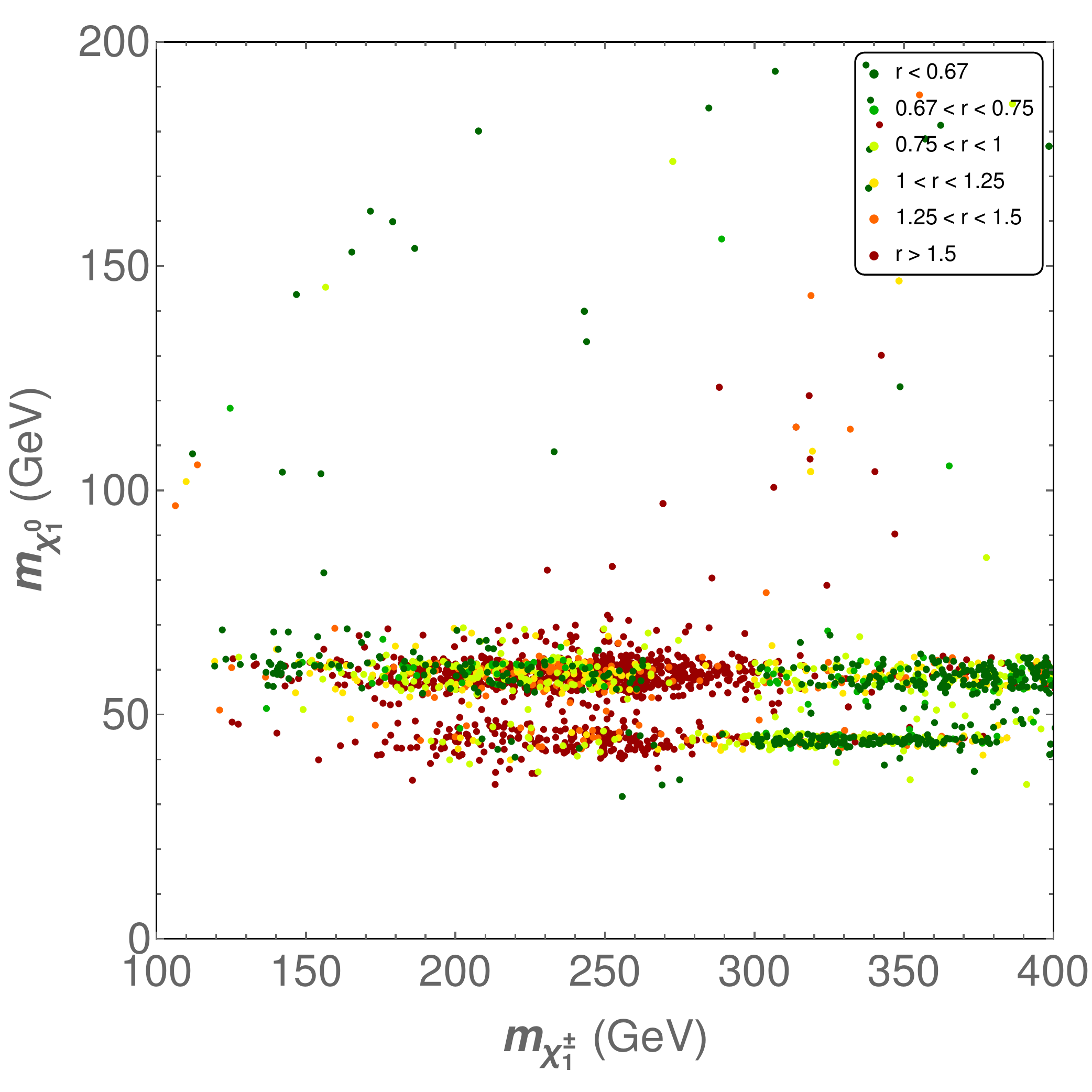} 
		\includegraphics[width=0.67\textwidth,scale=0.8]{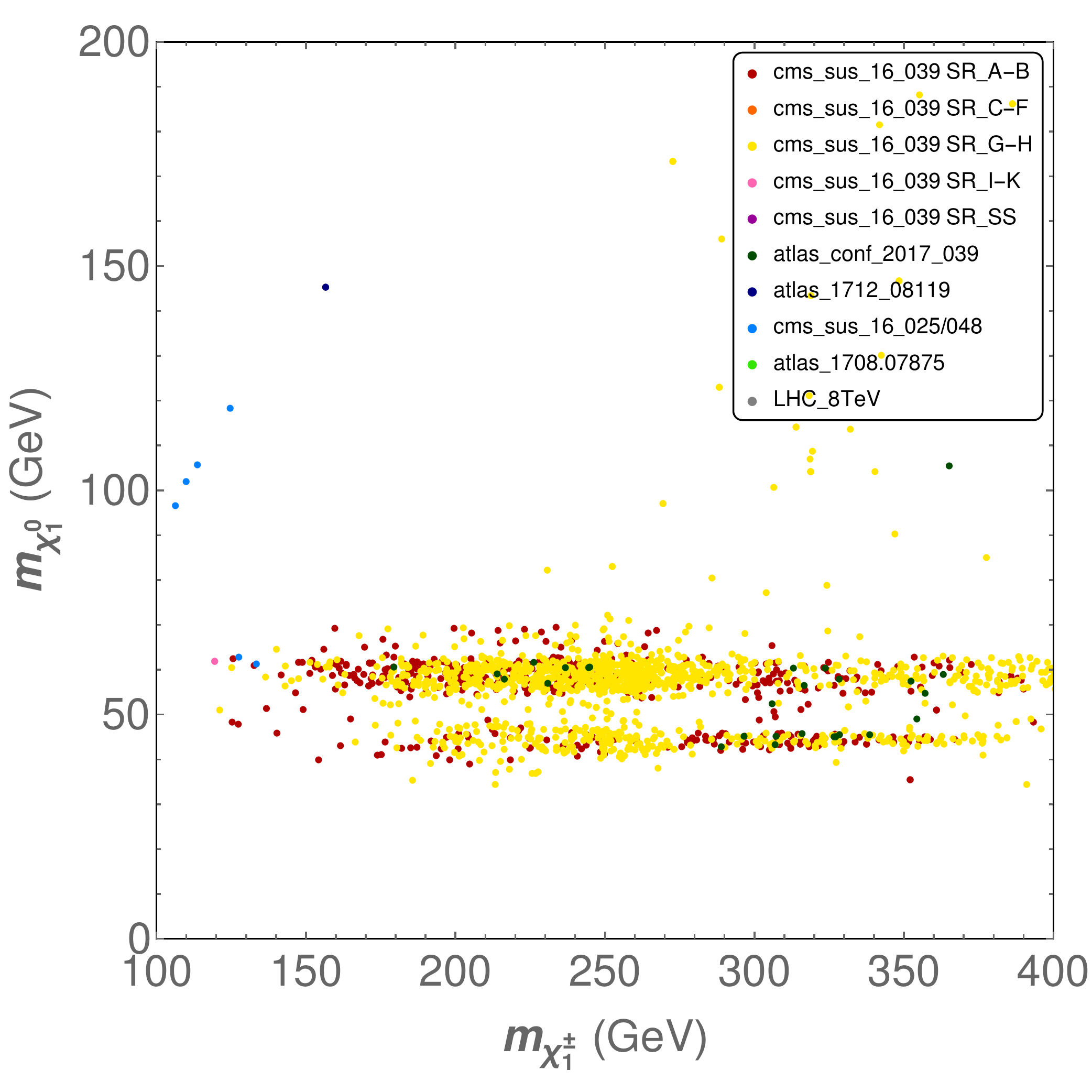}
		\caption{Model points in the $\chi_1^\pm$ and $\chi_1^0$ mass plane for the singlino NLSP scenario. These plots follow the color code of Fig.~\ref{fig:mssm}.}
		\label{fig:singlino_nlsp}
	\end{figure}
	
	In general, the existence of a singlino NLSP could have little to no impact on the collider phenomenology, if this state is ignored by the other, more efficiently produced electroweakino states. As discussed in Subsect.~\ref{subsec:scenarios}, we decided to focus on points with decays of at least 30\% of the heavier neutralino states (with mass below $300$~GeV) into the singlino NLSP. Naively, this condition could blurr the limits from $3\ell$-searches by opening the channel $\chi_3^0\to \Phi^{(*)}(\chi_2^0\to\Phi^{(*)}\chi_1^0)$, where $\Phi^{(*)}$ represents a possibly off-shell (singlet or doublet) Higgs or gauge boson. The consequences can take various forms. First, the final states are modified, possibly richer in light leptons (if $\Phi$ has a significant decay into light leptons) than in the MSSM-like scenario, or poorer (if $\Phi$ is a Higgs boson, with subsequent decays into heavy SM fermions, or a photon, due to reduced available phase-space between the electroweakino states). In addition, the leptons produced in this decay chain will tend to be less energetic than for a direct decay to the LSP, because of the presence of an intermediate step. This would result in a weaker efficiency of the searches.
	
	Fig.~\ref{fig:singlino_nlsp} presents the sensitivity of LHC multi-lepton searches in the same mass plane as previously. As in the MSSM-like scenario, the limits included within \texttt{CheckMATE} lead to the exclusion of many points in the funnel regions, with chargino masses up to $\sim400$~GeV. Interestingly, however, many spectra with $m_{\chi_1^{\pm}}\lsim200$~GeV appear to escape these constraints: there, the presence of a singlino NLSP together with light charginos leads to compressed spectra with soft leptonic final states, which are more challenging to access experimentally. Beyond the CMS searches for $3$ or more light leptons \cite{CMS-PAS-SUS-16-039}, we observe that the signal regions with more than $3$ leptons ({SR G}) are much more relevant than in the MSSM-like scenario. We can understand this as follows: the presence of an additional ladder (singlino NLSP) in the decay chain increases the multiplicity of leptons in the final state, leading to an increased relevance of the associated search channels. In fact, we can relate the higher density of excluded points in the chargino mass-range $\sim200-300$~GeV to the performance of the $4\ell$ searches. We also checked that low $2\ell$ and $3\ell$ cross sections were more frequently obtained in the funnel regions than for the MSSM-like scenario.
	
	We observe that many points with a chargino mass in the range $m_{\chi_1^\pm}\approx150\dots300$ GeV contain a large branching ratio of the photonic decay of the singlino NLSP $\chi_2^0\to\chi_1^0\gamma$. This is achieved in situations where the 2-body decays employing massive gauge or Higgs bosons are kinematically forbidden. Consequently, the final states are quite rich in photons, so that the search for (comparatively soft) photon final states could offer a viable alternative in view of probing this type of compressed spectra.
	
	\paragraph{Test-points --}
	Table~\ref{singlinoNLSPexample} collects a few benchmark points representative of this scenario.
	\begin{table}[t]
		\begin{center}
			\begin{tabular}{|c||c|c|c|c|c|}
				\hline
				& \verb|58_B9D|                  & \verb|20_B9E|                  & \verb|132_B9B|                 & \verb|62_B9D|                  & \verb|24_B9F|                  \\ \hline\hline
				$m_{\chi_1^0}$ (GeV)     & B\ \ \ \ $48$                  & B\ \ \ \ $62$                  & B\ \ \ \ $59$                  & B\ \ \ \ $59$                  & B\ \ \ \ $62$                  \\ \hline
				$m_{\chi_2^0}$ (GeV)     & S\ \ \ \ $82$                  & S\ \ \ \ $73$                  & S\ \ \ \ $75$                  & S\ \ \ \ $111$                 & S\ \ \ \ $175$                 \\ \hline
				$m_{\chi_1^{\pm}}$ (GeV) & W/H\ $127$                     & H\ \ \ $123$                   & H\ \ \ $139$                   & H\ \ \ $236$                   & W\ \ \ $241$                   \\ \hline\hline
				$m_{H_1^0}$ (GeV)        & $48$                           & $70$                           & $55$                           & $110$                          & $125$                          \\ \hline
				$m_{A_1^0}$ (GeV)        & $116$                          & $15$                           & $102$                          & $142$                          & $247$                          \\ \hline\hline
				BR$[\chi_1^{\pm}\to\chi_1^0W]$ & $1^*$                    & $0.81^*$                       & $0.92^*$                       & $0.30$                         & $0.45$                         \\ \hline
				BR$[\chi_i^0\to\chi_1^0Z]$ (pb) & $0.27^*_{i=3}$          & $0.01^*_{i=3}$; $0.18^*_{i=4}$ & $0.16_{i=3}$; $0.55_{i=4}$  & $0.08_{i=3}$; $0.23_{i=4}$     & $0.94_{i=3}$                   \\ \hline\hline
				$\sigma_{\text{8\,TeV}}[pp\to\chi_i^0\chi_1^{\pm}]$ (pb) & $1.80$ ($i=3$) & $0.28$ ($i=3+4$)               & $0.75$ ($i=3+4$)               & $0.12$ ($i=3+4$)               & $0.17$ ($i=3$)                 \\ \hline
				$\sigma_{\text{8\,TeV}}[pp\to\chi_1^+\chi_1^-]$ (pb) & $1.01$             & $0.11$                         & $0.31$                         & $0.06$                         & $0.09$                         \\ \hline
				$\sigma_{\text{13\,TeV}}[pp\to\chi_i^0\chi_1^{\pm}]$ (pb) & $3.66$ ($i=3$) & $0.37$ ($i=3+4$)               & $1.57$ ($i=3+4$)               & $0.28$ ($i=3+4$)               & $0.41$ ($i=3$)                 \\ \hline
				$\sigma_{\text{13\,TeV}}[pp\to\chi_1^+\chi_1^-]$ (pb) & $2.07$             & $1.12$                         & $0.65$                         & $0.13$                         & $0.21$                         \\ \hline\hline
				$\sigma_{\text{8\,TeV}}[pp\to3\ell]$ (fb)   & $24$                      & $2$                            & $9$                            & $3$                       & $4$                        \\ \hline
				$\sigma_{\text{13\,TeV}}[pp\to3\ell]$ (fb)   & $51$                            & $3$                            & $18$                            & $8$                            & $9$                            \\ \hline\hline
				Search                   & { SR A$02$}       & { SR0$\tau$a$16$}   & { SR A$19$}       & { SR G$05$}       & { SR A$30$}       \\ \hline
				$r$                      & $1.2$                          & $0.3$                          & $0.2$                          & $2.4$                          & $3.3$                          \\ \hline
			\end{tabular}
			\caption{Test-points for the singlino NLSP scenario. The general features are similar to those of Table~\ref{MSSMexample}. \label{singlinoNLSPexample}}
		\end{center}
		\vspace{-2ex}
	\end{table}
	\begin{itemize}
		\item \verb|58_B9D|: The spectrum of this point is reminiscent of that of \verb|39_A18| in the MSSM-like scenario: the bino LSP with mass $\sim48$~GeV is in the $Z$-funnel; mostly-wino states take a mass of $\sim130$~GeV, while mostly-higgsino states take a mass of $\sim270$~GeV -- winos and higgsinos are sizably mixed, though. In addition, a singlino NLSP intervenes at a mass of $\sim82$~GeV. The decays of this state are essentially mediated by an off-shell $Z$ and thus appear relatively conventional (although $BR[\chi_2^0\to\chi_1^0\gamma\simeq8\%]$). The decays of $\chi_3^0$ (resp.~$\chi_4^0$) involve the singlino-like state at $\sim40\%$ (resp.~$\sim25\%$). While the leptonic cross sections at the LHC, dominated by the $\chi_{3,4}^0\chi_1^{\pm}$ channels, are comparable to those obtained for \verb|39_A18|, an appreciable proportion of the produced leptons originate in the decays of $\chi_2^0$ and are thus relatively soft. Therefore, these leptons are identified with only a weak efficiency. According to \texttt{CheckMATE}, Run-1 searches are insensitive to this point. At Run-2, it is possible to constrain this spectrum via $3\ell$ searches, though $r$ remains below $1.5$. The most sensitive signal region is SRA02 with a cut on the invariant dilepton mass below the $Z$-mass window as well as a weak cut on missing transverse momentum and $m_T\le100$ GeV. This search is thus able to access the leptons originating in the subleading $\chi_3^0\to\chi_1^0 Z^*$ decay. We expect a clear exclusion to be within reach of a somewhat larger integrated luminosity.
		\item \verb|20_B9E| involves a bino-like LSP in the SM-Higgs funnel. The singlino is about $\sim10$~GeV heavier. The higgsinos and winos take mass in the range of $\sim120$ and $300$~GeV, respectively. The production cross sections of the higgsino states in $p-p$ collisions are comparatively small. In addition, the higgsinos have sizable decays into the light CP-odd singlet Higgs and the singlino. The latter mostly decays via photonic or hadronic channels. The resulting signals in the multi-lepton channels are considerably suppressed, leading to a good agreement with the experimental limits. \texttt{CheckMATE} identifies a $3\ell$ search of Run-1 as the most `sensitive' channel.
		\item \verb|132_B9B| contains a bino LSP in the SM-Higgs funnel, a singlino NLSP at $\sim75$~GeV, higgsino states at $\sim140$~GeV and winos at $\sim850$~GeV. The production cross section of the higgsinos are larger than for the previous point. The decays of these states largely involve the singlino and the CP-even Higgs singlet at $\sim55$~GeV. Then, the singlino has sizable photonic decays, while the singlet Higgs essentially decays into $b\bar{b}$ pairs. Again, \texttt{CheckMATE} concludes to little sensitivity of the multi-lepton searches.
		\item \verb|62_B9D|: The spectrum contains a bino-like LSP with mass $\sim59$~GeV, a singlino-like NLSP (at $\sim97\%$) with mass $\sim111$~GeV, higgsino states at $\sim240$~GeV and winos at $\sim400$~GeV. The production cross section of electroweakinos is relatively small due to the comparatively large mass of the winos and higgsinos, so that the $3\ell$ cross sections at the LHC are modest. The decays of the higgsinos and winos involve $\chi_2^0$ (at $\geq70\%$) while the main decay channel of the NLSP (at $91\%$) is $\chi_2^0\to\chi_1^0\gamma$ (its dominant `active' component is higgsino-like at $\sim2\%$). This induces final states that are rich with photons. This suggests that photonic searches could be employed to probe this type of spectrum. However, the results from Run-2 in the multi-lepton channels (more than $3\ell$) are able to exclude this point.
		\item \verb|24_B9F|: This point features a bino-like LSP at $\sim62$~GeV and a singlino-like NLSP at $\sim175$~GeV. The heavier electroweakinos are dominantly winos (with mass $\sim240$~GeV) and higgsinos (with mass $\sim320$~GeV). Except for the presence of the singlino, this point is comparable to \verb|spectr31| of 
		the MSSM-like scenario. The singlino copiously intervenes in the decays of the mostly-higgsino states (over $50\%$), but subdominantly in the decays of the lighter winos ($<1\%$). The latter are the main contributors to the leptonic cross sections in $p-p$ collisions. In addition, the singlino decays are conventional and mostly proceed through an on-shell $Z$. The large mass gap between LSP and NLSP also ensures energetic leptons in the final state. Thus, the additional ladder due to the presence of a singlino does not particularly endanger the traditional search strategy. According to \texttt{CheckMATE}, Run-1 results already hint at tensions in the $3\ell$ searches. This is confirmed by the CMS searches at $13$~TeV.
	\end{itemize}
	
	\subsection{Decays into Higgs singlets}
	Light singlet-like Higgs states are phenomenologically realistic in the NMSSM. In the scenario with light singlino LSP, the annihilation of DM in the early Universe is often mediated by such a Higgs boson. However, light singlet Higgs states could also exist irrespectively of the presence of a very light DM candidate. In this subsection, we investigate the impact of the LHC searches on this type of scenario.
	
	\paragraph{Global analysis --} As in the scenario with singlino NLSP, the produced electroweakinos do not necessarily involve the singlet state in their cascade decays and, in such a case, the electroweakino phenomenology largely reduces to that of a MSSM-like point. Therefore, the spectra of our sample satisfy the additional requirement that the Higgs singlet intervenes at $\gsim10\%$ in the decays of the NLSP. The presence of this extra Higgs state in the decay chain is expected to 
	increase the proportion of $b\bar{b}$ or $\tau\tau$ final states, hence reducing the effectiveness of lepton final state searches since only hadronic taus (BR$\sim65\%$) can be tagged while the efficiency of hadronic taus ($\epsilon_\tau\sim40\sim70\%$) is much worse than that of light leptons. In this sample, most points possess a bino LSP and cluster in the $Z$-\ or SM Higgs-funnels. Wino or higgsino LSP's are also represented in the coannihilation region. Finally, occasional points (with mostly bino LSP) may annihilate in the Higgs-singlet funnel. Such points typically broaden the $Z$ / SM-Higgs funnels. We will consider this latter case more closely in the following subsection.
	
	\begin{figure} \centering
		\includegraphics[width=0.67\textwidth,scale=0.8]{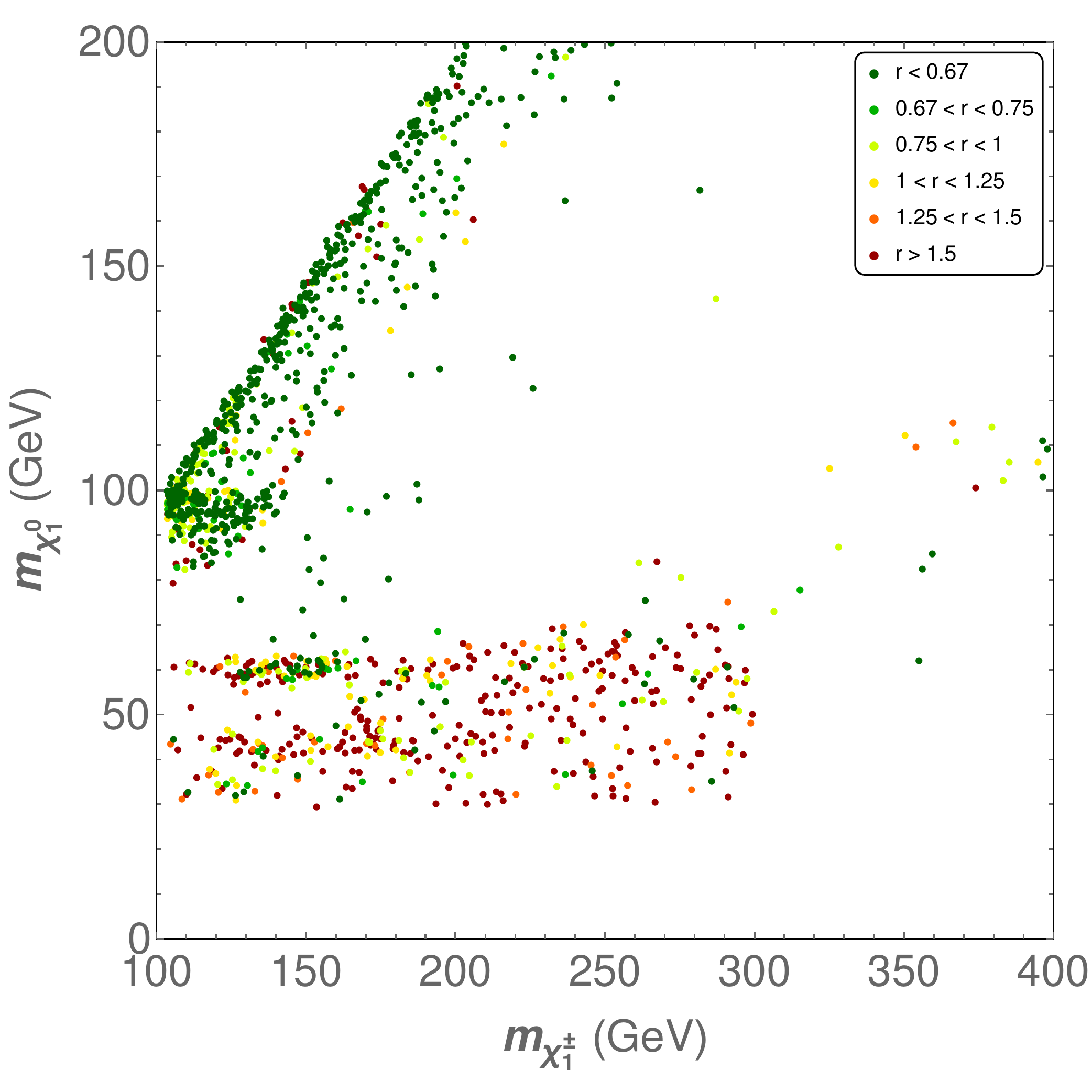} \includegraphics[width=0.67\textwidth,scale=0.8]{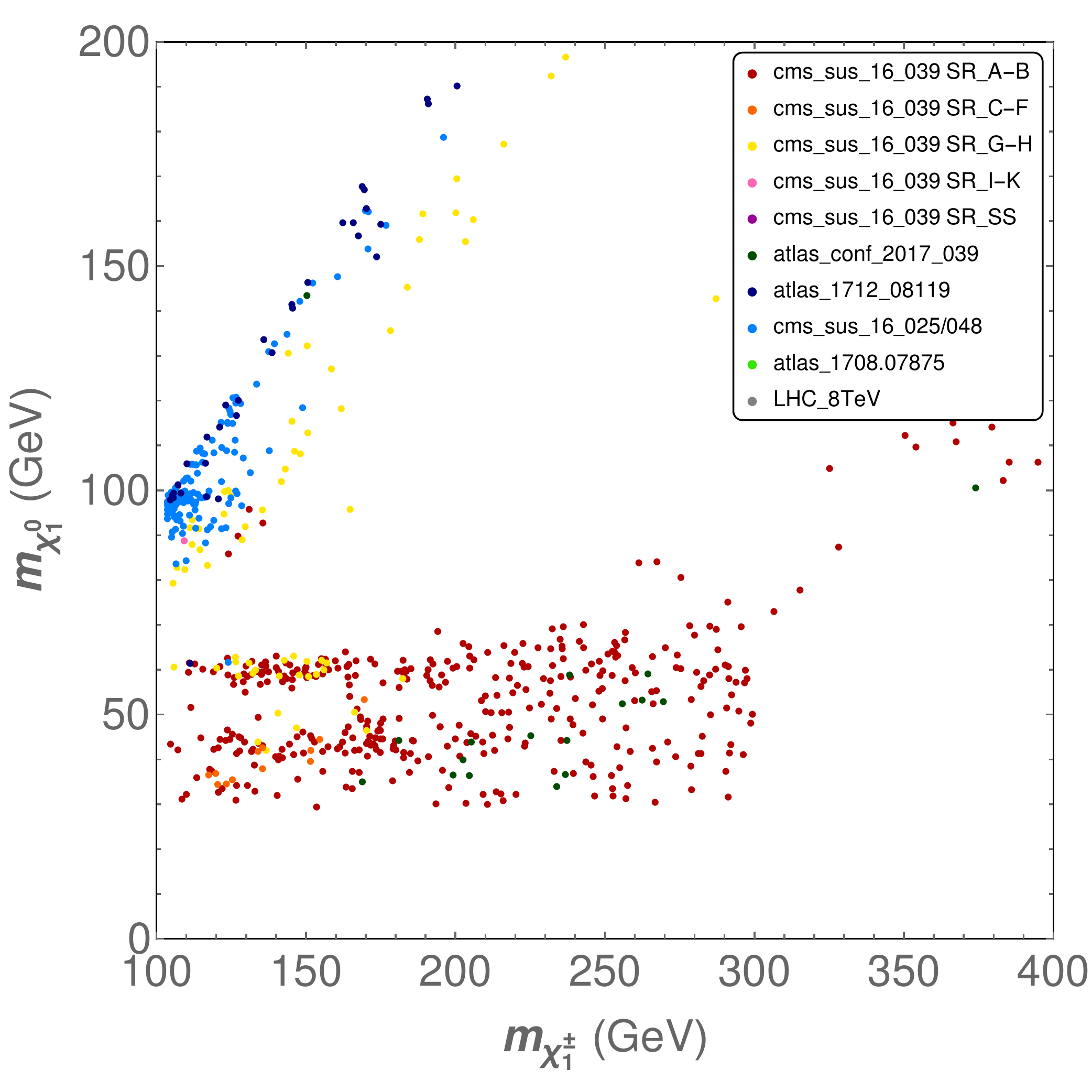}  
		\caption{Model points in the $\chi_1^\pm$ and $\chi_1^0$ mass plane for the scenario with light Higgs singlets in electroweakino decays. These plots follow the color code of Fig.~\ref{fig:mssm}. \label{fig:NMSSM_light_singlet}}
	\end{figure}
	
	In Fig.~\ref{fig:NMSSM_light_singlet}, we present the limits resulting from LHC searches. The general aspect is comparable to that obtained in the MSSM-like scenario. However, the coannihilation region appears to be less efficiently probed than in the MSSM-like case and we observe a higher density of allowed points at the base of the funnel region. In fact, we checked that, while the maximal values of the $2\ell$ and $3\ell$ cross sections mediated by the electroweakinos are comparable to those of the MSSM-like points, much lower values are also accessible. Indeed, the opening of decays through a light Higgs singlet is expected to reduce the relevance of light-leptonic final states. 
	
	Once again, the \texttt{CheckMATE} analysis identifies the CMS searches at a center-of-mass energy of $13$~TeV, with $3$ or more leptons in the final state (type {SR A} and {SR G}), as the most constraining signal regions. We identify a marginal sensitivity of searches with $\tau$ final states (type SR~F) for points with low chargino masses (large production cross-sections). In practice, no clear exclusion is obtained for any of these spectra with the preliminary Run-2 data.
	
	\paragraph{Test-points --}
	We already encountered a few points involving light singlet states mediating the electroweakino decays, e.g.\ in the scenario with singlino NLSP (\verb|20_B9E| for example). A few additional points that are representative of this scenario are provided in Table~\ref{lightsingletexample}.
	
	\begin{table}[t]
		\begin{center}
			\begin{tabular}{|c||c|c|c|c|c|c|}
				\hline
				& \verb|223_C1B_C1|              & \verb|30_C1B_C1|               & \verb|3_C1B_C19A|              & \verb|4_C1A_C21|                & \verb|47_C1B_C1|                \\ \hline\hline 
				$m_{\chi_1^0}$ (GeV)     & B\ \ \ \ $45$                  & B\ \ \ \ $44$                  & B\ \ \ \ $60$                  & B\ \ \ \ $68$              & B\ \ \ \ $35$                  \\ \hline       
				$m_{\chi_2^0}$ (GeV)     & W\ \ \ $127$                   & W\ \ \ $223$                   & H\ \ \ $153$                   & H\ \ \ $169$               & W\ \ \ $120$                    \\ \hline       
				$m_{\chi_1^{\pm}}$ (GeV) & W\ \ \ $128$                   & W\ \ \ $223$                   & H\ \ \ $147$                   & H\ \ \ $152$               & W\ \ \ $120$                    \\ \hline\hline 
				$m_{H_1^0}$ (GeV)        & $125$                          & $123$                          & $125$                          & $64$                       & $127$                            \\ \hline       
				$m_{A_1^0}$ (GeV)        & $74$                           & $65$                           & $26$                           & $178$                      & $71$                           \\ \hline\hline 
				BR$[\chi_1^{\pm}\to\chi_1^0W]$ & $1$                      & $1$                            & $1$                            & $1$                        & $1$                             \\ \hline       
				BR$[\chi_i^0\to\chi_1^0Z]$  & $0.87^*$ ($i=2$)            & $0.88$ ($i=2$)                 & $0.88_{i=2}$; $0.61_{i=3}$     & $0.93_{i=2}$; $0.05_{i=3}$ & $0.17^*_{i=2}$      \\ \hline\hline 
				$\sigma_{\text{8\,TeV}}[pp\to\chi_i^0\chi_1^{\pm}]$ (pb) & $2.36$ ($i=2$) & $0.27$ ($i=2$)                 & $0.55$ ($i=2+3$)               & $0.44$ ($i=2+3$)                & $3.4$ ($i=2$)                \\ \hline       
				$\sigma_{\text{8\,TeV}}[pp\to\chi_1^+\chi_1^-]$ (pb) & $1.37$             & $0.15$                         & $0.25$                         & $0.23$                          & $1.7$                          \\ \hline 
				$\sigma_{\text{13\,TeV}}[pp\to\chi_i^0\chi_1^{\pm}]$ (pb) & $4.78$ ($i=2$) & $0.63$ ($i=2$)                 & $1.16$ ($i=2+3$)               & $0.92$ ($i=2+3$)                & $6.7$ ($i=2$)                \\ \hline       
				$\sigma_{\text{13\,TeV}}[pp\to\chi_1^+\chi_1^-]$ (pb) & $2.80$             & $0.37$                         & $0.53$                         & $0.49$                         & $3.4$                         \\ \hline\hline 
				$\sigma_{\text{8\,TeV}}[pp\to3\ell]$ (fb)   & $33$                           & $4$                           & $7$                           & $5$                            & $10$                            \\ \hline       
				$\sigma_{\text{13\,TeV}}[pp\to3\ell]$ (fb)   & $68$                            & $10$                            & $14$                            & $11$                             & $21$                             \\ \hline\hline       
				Search                   & { SR A$08$}       & { SR A$30$}       & { SR A$25$}       & { SR A$24$}        & { SR F$02$}        \\ \hline       
				$r$                      & $4.7$                          & $3.1$                          & $0.6$                          & $0.4$                    & $0.96$                           \\ \hline       
			\end{tabular}
			\caption{Test-points for the scenario with light Higgs singlets. The general features are similar to those of Table~\ref{MSSMexample}. \label{lightsingletexample}}
		\end{center}
		\vspace{-2ex}
	\end{table}
	\begin{itemize}
		\item \verb|223_C1B_C1|: The electroweakino spectrum contains a bino-like LSP with mass of $\sim45$~GeV ($Z$-funnel), wino-like states at $\sim127$~GeV and higgsino-like states at $\sim300$~GeV. This is comparable to the characteristics of the point \verb|39_A18| of the MSSM-like scenario. Similarly, the production cross section of the wino-like states in $p-p$ collisions reaches a few pb and leads to sizable lepton+MET cross sections. The presence of a light singlet-like pseudoscalar with mass $\sim74$~GeV has little impact on the decay chains. Only a small fraction of the neutralino decays are mediated by the singlet scalar $A_1$: almost $87\%$ of the neutralino proceed through an off-shell $Z$-boson, so that multilepton final states are quite common. \texttt{CheckMATE} already identifies tensions with the $3\ell$ results of Run-1. Exclusion is 	achieved at Run-2 in the same channel. SR\,A08 is the most sensitive signal region, with cuts optimized for invariant-mass pairs below the $Z$-mass window and a transverse cut above the $M_W$ endpoint.
		This illustrates the fact that the presence of a light singlet like state does not automatically yield significant effects on the phenomenology of electroweakino states at colliders.
		\item \verb|30_C1B_C1|: The spectrum is similar to that of \verb|spectr31| of the MSSM-like scenario but, in addition, a CP-odd singlet-like Higgs is present at a mass of $\sim65$~GeV. Chargino pair and associated charged and neutral wino final states are the prominent production channels with BR$(\chi_2^0\rightarrow\chi_1^0 A_1)=12.3\%$. Thus the presence of the light singlet pseudoscalar does not yield a deep impact on the phenomenology. Both the ATLAS and CMS multi-lepton searches do not have sensitivity to the chargino production channel. However, they both perform efficiently for the $\chi_1^\pm\chi_2^0$ channel. Tensions appear in the $3\ell$ searches of Run-1. The CMS search at $13$~TeV confirms the exclusion of this benchmark in the trilepton final state.
		\item \verb|3_C1B_C19A| contains a bino-like LSP in the SM-Higgs funnel, with higgsino NLSP at $\sim150$~GeV, singlino/wino states at $\sim800$~GeV. Higgsino production in $p-p$ collisions at $8$~TeV amount to a few $100$~fb. The decays of these states involve the light CP-odd singlet of $\sim26$~GeV at the level of $12\%$ ($\chi_2^0$)	and $40\%$ ($\chi_3^0$). In comparison to a $Z$-boson, the presence of the light Higgs in the decay chain increases the probability of a $\tau$-pair with low invariant squared mass, while final states with only light leptons are less frequent. Chargino pair production is also quite frequent however: the mass-splitting between the chargino and LSP is roughly the $W$ mass and the events closely resemble the SM $WZ$ background. Thus, this final state is difficult to probe. According to \texttt{CheckMATE}, both the Run-1 and Run-2 multi-lepton searches are blind to this spectrum. 
		\item \verb|4_C1A_C21|: For this point, the light singlet Higgs is CP-even and has a mass of $\sim64$~GeV. It enters the decays of the neutral higgsinos at $\sim170$~GeV. Singlino and winos are substantially heavier ($\sim500$~GeV). The $\chi_1^\pm\chi_{2,3}^0$ channel would potentially deliver the best sensitivity. However, a sizable fraction of the higgsino decays employ the Higgs channels, reducing the relevance of light leptons in the final state. In addition, although $\chi_2^0$ has sizable decays into the $Z$ boson, its production cross section is reduced (due to the higgsino nature) and the events associated to $\chi_1^\pm\chi_{2}^0$ look $WZ$-like, making it difficult to separate the signal events from the SM background events. No constraints from multi-lepton searches apply.
		\item \verb|47_C1B_C1|: In this case, the bino LSP at a mass of $\sim35$~GeV can annihilate through the funnel of a light CP-odd Higgs at $\sim71$~GeV. The wino NLSP states have a mass of $\sim120$~GeV, leading to sizable production cross-sections both at Run-1 and Run-2. The decays of the neutralino $\chi_2^0$ dominantly employ the $\chi_1^0A_1$ channel (at $\sim83\%$). $8$~TeV searches appear to be completely blind to this point. At $13$~TeV, the signal regions with light leptons offer a marginal sensitivity to the spectrum and the CMS searches of F-type (one light lepton and a tau pair) prove competitive. However, the sensitivity remains as yet too loose to conclude to a clear exclusion of this spectrum.
	\end{itemize}
	
	\subsection{Higgs singlet on LSP annihilation threshold}
\begin{figure} \centering
		\includegraphics[width=0.67\textwidth,scale=0.8]{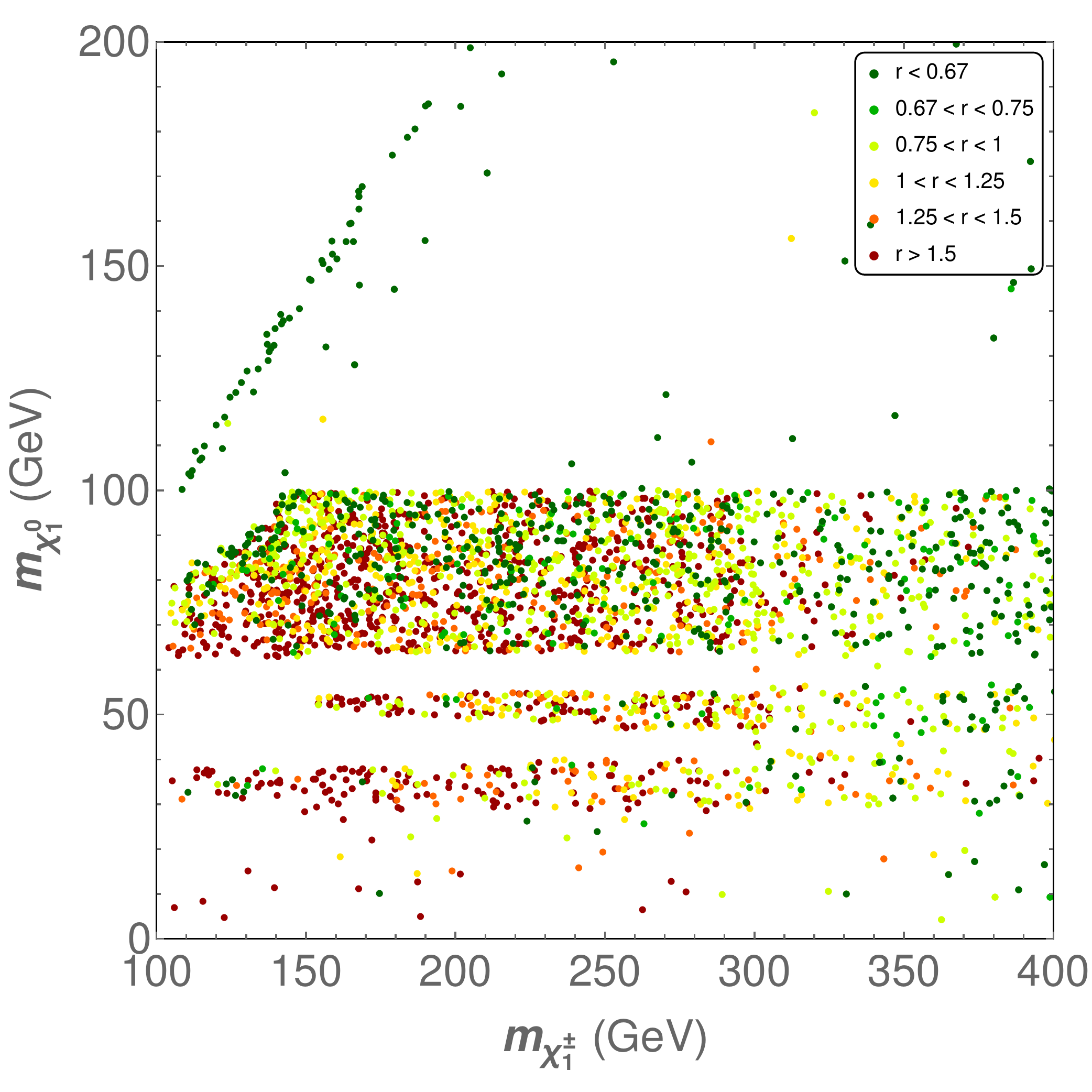} 
		\includegraphics[width=0.67\textwidth,scale=0.8]{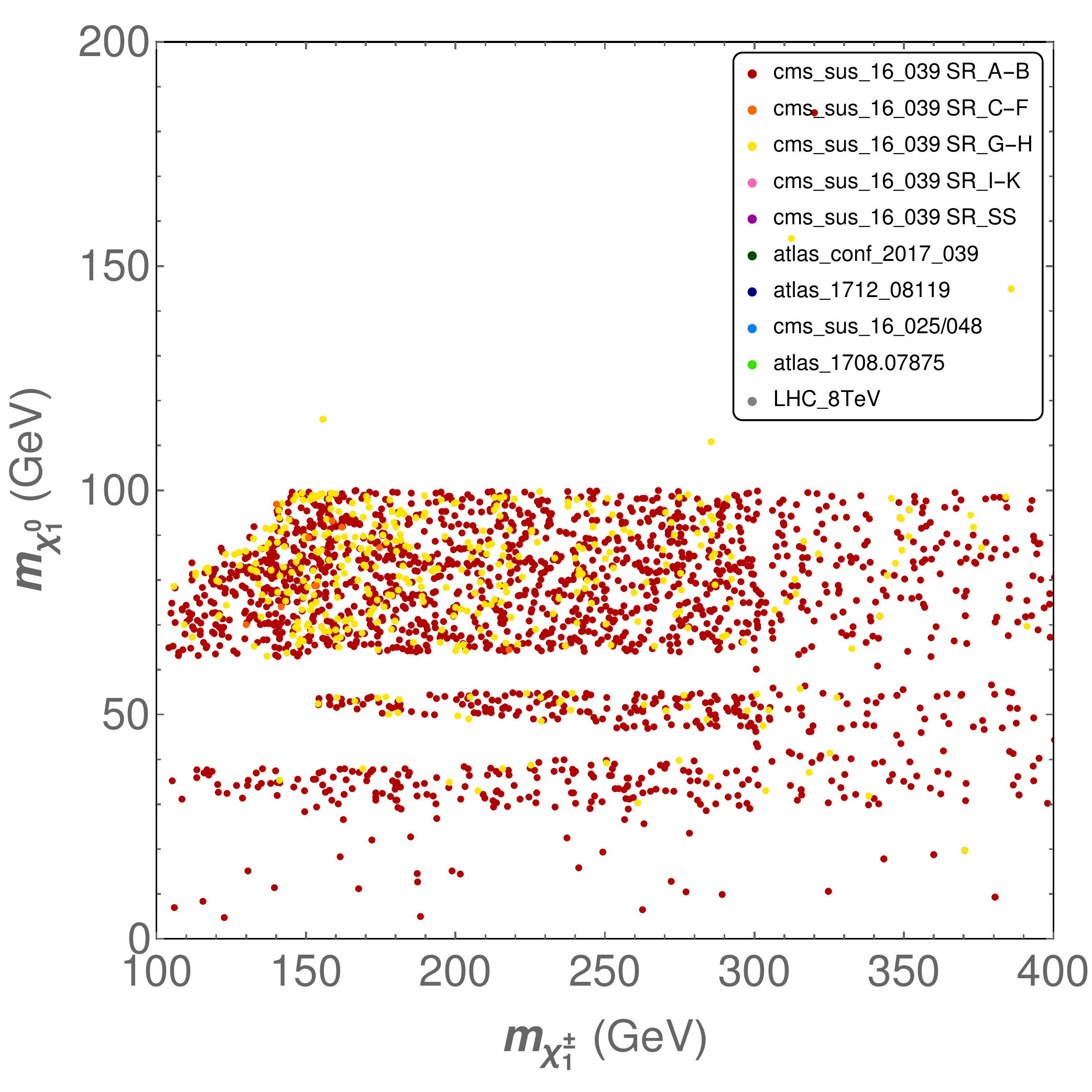} 
		\caption{Model points in the $\chi_1^\pm$ and $\chi_1^0$ mass plane for the Higgs singlet on LSP in annihilation threshold. These plots follow the color code of Fig.~\ref{fig:mssm}.}
		\label{fig:NMSSM_singlet_th}
	\end{figure}

	Finally, we consider a scenario where the relic density constraint is satisfied via the mediation of a resonant singlet Higgs state in DM annihilation. The mass of the (CP-even or odd) singlet is thus approximately twice the LSP mass. We already encountered this type of mediation in the context of light singlino LSP. However, the 
	large majority of the points populating the current sample involve a bino LSP, with occasional higgsino/winos at the fringe with $m_{\chi_1^0}\approx m_{\chi_1^{\pm}}$. In order to characterize the singlet-mediation scenario more closely, we decided to exclude $Z$- and SM-Higgs funnels as well as the coannihilation region. This is reflected in Fig.~\ref{fig:NMSSM_singlet_th} through the unpopulated regions in the LSP--chargino mass plane corresponding to the excluded parameter space of the scan. The sparsity of points for $m_{\chi_1^0}>100$~GeV is an artifact of the scan, as we concentrated the numerical effort on $m_{\chi_1^0}<100$~GeV.
	
	\paragraph{Global Analysis --}
	Again, we observe a very inhomogeneous pattern of constraints in the plane defined by the LSP and $\chi_1^{\pm}$ masses. Excluded points seem to appear with a larger density for lighter LSP's and charginos, while points with a chargino mass above $\sim300$~GeV are generally allowed. In fact, the pattern of exclusion essentially extends outside of the $Z$/$H_{\text{SM}}$-funnel and coannihilation regions the features observed in the MSSM-like scenario. Most of the time, the light singlet Higgs only intervenes in the relic-density condition, without having a particular effect on the collider phenomenology (contrarily to the points targeted in the previous scenario). Thus, the collider phenomenology is essentially that of a bino-like LSP and remains largely unchanged with respect to the MSSM-like case (up to the access to a wider range of kinematical configurations). Correspondingly, the CMS $3\ell$ search at $13$~TeV center-of-mass energy again emerges as the most efficient of the multi-lepton searches.

	\paragraph{Test-points --}
	Table~\ref{thressingletexample} provides a few examples of points with LSP annihilation in a light-singlet Higgs funnel.
	\begin{table}[t]
		\begin{center}
			\begin{tabular}{|c||c|c|c|c|c|c|}
				\hline
				& \verb|122_C2B_C1|              & \verb|543_C2B_C1|              & \verb|203_C2B_C15|             & \verb|186_C2B_C15|            & \verb|195_C2B_C1|               \\ \hline\hline
				$m_{\chi_1^0}$ (GeV)     & B\ \ \ \ $38$                  & B\ \ \ \ $30$                  & B\ \ \ \ $5$                   & B\ \ \ \ $36$                 & B\ \ \ \ $70$                   \\ \hline
				$m_{\chi_2^0}$ (GeV)     & W\ \ \ $116$                   & W\ \ \ $227$                   & W\ \ \ $188$                   & W\ \ \ $125$                  & W\ \ \ $113$                    \\ \hline
				$m_{\chi_1^{\pm}}$ (GeV) & W\ \ \ $117$                   & W\ \ \ $227$                   & W\ \ \ $188$                   & W\ \ \ $125$                  & W\ \ \ $113$                    \\ \hline\hline
				$m_{H_1^0}$ (GeV)        & $125$                          & $125$                          & $126$                          & $123$                         & $126$                           \\ \hline
				$m_{A_1^0}$ (GeV)        & $80$                           & $64$                           & $10$                           & $73$                          & $145$                           \\ \hline\hline
				BR$[\chi_1^{\pm}\to\chi_1^0W]$ & $1^*$                    & $1$                            & $1$                            & $1$                           & $1$                             \\ \hline
				BR$[\chi_i^0\to\chi_1^0Z]$ & $0.81^*$ ($i=2$)             & $0.72$ ($i=2$)                 & $0.54$ ($i=2$)                 & $0.07^*$ ($i=2$)              & $0.97^*$ ($i=2$)                \\ \hline\hline
				$\sigma_{\text{8\,TeV}}[pp\to\chi_i^0\chi_1^{\pm}]$ (pb) & $2.69$ ($i=2$) & $0.29$ ($i=2$)                 & $0.48$ ($i=2$)                 & $0.22$ ($i=2$)                & $4.67$ ($i=2$)                  \\ \hline
				$\sigma_{\text{8\,TeV}}[pp\to\chi_1^+\chi_1^-]$ (pb) & $1.41$             & $0.16$                         & $0.24$                         & $1.60$                        & $2.36$                          \\ \hline
				$\sigma_{\text{13\,TeV}}[pp\to\chi_i^0\chi_1^{\pm}]$ (pb) & $5.36$ ($i=2$) & $0.67$ ($i=2$)                 & $1.07$ ($i=2$)                 & $0.44$ ($i=2$)                & $9.28$ ($i=2$)                  \\ \hline
				$\sigma_{\text{13\,TeV}}[pp\to\chi_1^+\chi_1^-]$ (pb) & $2.87$             & $0.37$                         & $0.55$                         & $3.28$                        & $4.73$                          \\ \hline\hline
				$\sigma_{\text{8\,TeV}}[pp\to3\ell]$ (fb)   & $48$                    & $3$                      & $7$                           & $0$                           & $32$                             \\ \hline
				$\sigma_{\text{13\,TeV}}[pp\to3\ell]$ (fb)   & $100$                            & $8$                            & $18$                            & $1$                           & $63$                             \\ \hline\hline
				Search                   & { SR A$01$}       & { SR A$30$}       & { SR A$27$}       & { WWa SF}        & { SR A$01$}        \\ \hline
				$r$                      & $3.4$                          & $2.1$                          & $1.7$                          & $0.5$                         & $0.6$                           \\ \hline
			\end{tabular}
			\caption{Test-points for the scenario where Higgs singlets mediate the DM annihilation. The general features are similar to those of Table~\ref{MSSMexample}. \label{thressingletexample}}
		\end{center}
		\vspace{-2ex}
	\end{table}
	\begin{itemize}
		\item \verb|122_C2B_C1|: The spectrum includes a bino-like LSP at $\sim38$~GeV that annihilates in the CP-odd singlet Higgs funnel ($m_{A_1^0}\simeq80$~GeV), wino states at $\sim115$~GeV and higgsino states at $\sim225$~GeV. The singlino is very heavy ($\sim5$~TeV). As the electroweakino sector is light, the production cross section of the mostly-wino states ($\chi_2^0\chi_1^{\pm}$, $\chi_1^+\chi_1^-$) is rather large. The decays of these states is then essentially mediated by gauge bosons (the light pseudoscalar does not intervene in the decay chain), leading to sizable lepton cross sections. As $m_{\chi_2^0}-m_{\chi_1^0}>70$~GeV, the leptons are relatively hard, so that the lepton searches at the LHC prove efficient: according to \texttt{CheckMATE}, the ATLAS $3$-lepton search of Run-1 is able to exclude the point, with also tensions in the $2$-lepton {WWa SF} signal region. This picture is confirmed by the $13$~TeV results. The most sensitive signal region targets events below the $Z$-mass window and applies the weakest cuts on the missing transverse momentum and the transverse invariant mass. Except for the LSP being off the funnel regions (in fact: on the CP-odd singlet funnel), the phenomenology of the electroweakino sector of \verb|122_C2B_C1| at colliders is essentially the same as for a MSSM-like scenario (e.g.\ \verb|39_A18|).
		\item \verb|543_C2B_C1|: Except for the bino-like LSP annihilating in the CP-odd singlet funnel, this point has similarities with \verb|spectr31| of the MSSM-like scenario. The mostly wino states have a mass of $\sim225$~GeV, while the higgsinos have a mass of $\sim600$~GeV and the singlino is very heavy ($\sim12$~TeV).
		The decays of the winos and higgsinos are essentially mediated by the gauge bosons -- the CP-odd Higgs has only a limited impact on the decay chains, intervening at $\sim20\%$ in the decays of neutralinos. Due to the relatively high mass, the production of winos returns moderate cross sections. However, the leptons are rather hard and thus efficiently detected by the experiment. While only a mild excess appears in Run-1 $3\ell$ searches, \texttt{CheckMATE} returns the exclusion of the point at $13$~TeV.
		\item \verb|203_C2B_C15|: This spectrum contains a light bino-like LSP with mass $\sim5$~GeV. Its annihilation is mediated by a singlet-like CP-odd Higgs with mass $\sim10$~GeV. The winos and higgsinos have masses of $\sim190$~GeV and $\sim320$~GeV. Half of the decays of the second lightest neutralino 
		involve the SM-like Higgs, while the decay into the CP-odd scalar at $\sim10$~GeV has a negligible impact on the decays of neutralinos and charginos. While the BR into $Z$ bosons represents only 
		50\% of the decays of $\chi_2^0$, the benchmark point is clearly excluded by the CMS trilepton signal region.
		\item \verb|186_C2B_C15|: This spectrum is similar to that of \verb|122_C2B_C1|. However, the production cross section of the wino-states $\chi_2^0\chi_1^+$ is notably suppressed, leading to a small $3\ell$ cross section. The dominant production channel is that of a chargino pair. Neither the ATLAS nor the CMS dilepton signal regions are sensitive to the former production channel. Moreover, we note that $\chi_2^0\to\chi_1^0 A_1$ is the main decay channel of the wino, placing this benchmark point beyond the reach of the current trilepton searches.
		\item \verb|195_C2B_C1|: Again, this point includes a relatively light bino/wino spectrum that appears to be unconstrained by multi-lepton searches. In this case, the production cross section of the winos is rather high and leads to sizable multilepton cross sections. The CP-odd singlet has a negligible impact on the electroweakino decays. Instead, the rather low mass gap $m_{\chi_2^0}-m_{\chi_1^0}<45$~GeV suggests that the leptons in the final state are too soft in view of the applied cuts.
	\end{itemize}
	
	\section{Outlook for the LHC-HL phase and alternative search channels}
	\paragraph{LHC-HL prospects --}
	Finally, we wish to conclude the discussion of the collider constraints by revisiting the discovery prospects of our electroweakino benchmark points for the future high-luminosity run. Our numerical analysis is based on the official ATLAS LHC-High Luminosity (HL) analysis \cite{ATL-PHYS-PUB-2014-010} that has been employed in a high-luminosity natural 
	SUSY study \cite{Kim:2016rsd}. The ATLAS study considers the direct electroweak production of charginos and neutralinos with decays via the SM gauge bosons and the SM Higgs at a center-of-mass energy of $\sqrt{s}=14$ TeV, together with an integrated luminosity of 3000 fb$^{-1}$. They take into account the configuration of the LHC-LH ATLAS detector, optimized selection cuts for the signal regions during the HL phase and, most importantly, the MC-derived estimation of background processes. The signal regions are optimized for the identification of $WZ$ and $WH$ final states. Again, the $WZ$ signal regions target trilepton final states with missing transverse momentum. A dilepton pair in the $Z$-mass window is requested, as well as a $b$-jet veto and a minimal cut on the transverse momentum for all three leptons of $p_T\ge50$ GeV. Four signal regions with cuts of increasing strength on the transverse mass and missing transverse momentum are defined. In addition, the $WH$ signal regions distinguish among a three-lepton and a hadronic-tau final states, with a strict $b$-jet veto in both cases. The former signal category focuses on searches for a SM Higgs decaying into two leptons via intermediate $ZZ$, $WW$ or $\tau\tau$. Here, SFOS lepton pairs are discarded and further cuts on the missing transverse momentum and the transverse masses of all involved light leptons are requested. The second topology targets hadronic taus originating from the SM Higgs decays. In this case, the invariant mass of the hadronic tau pair is requested to fall in the SM-Higgs mass window $80 \le m_{\tau\tau}\le130$ GeV.
	
	We tested all the test-points that continue to be allowed with the 8~TeV or the 13~TeV searches with $36$ fb$^{-1}$ against these high-luminosity prospects. The interesting but expected result is that the official ATLAS LHC-HL does not show any sensitivity to these benchmark points. On the one hand, the ATLAS study is not optimized to detect light electroweakinos with $m_{\chi_1^\pm}\le 200$ GeV. On the other hand, no signal region with light singlet scalars is considered, so that decays of singlet scalars into $b\bar b$ are systematically discarded by the $b$-jet veto. Light singlet decays into hadronic taus are also missed in general, since the hadronic $\tau^+\tau^-$ pairs are required to satisfy the cut selecting the SM-Higgs mass window. Finally, the ATLAS high-luminosity study does not consider the compressed region, which implies the absence of sensitivity in the co-annihilation region. 
	
	\paragraph{Alternative search channels --} 
	The numerical results clearly show that the multilepton -- and in particular the trilepton and large missing transverse momentum -- searches provides an effective coverage of the MSSM as well as NMSSM parameter space. However, as already in the MSSM, the NMSSM parameter regions cannot be fully excluded. The reasons are manifold, e.g.\ (i) the reduced cross sections for electroweakinos with higgsino/singlino admixtures compared to that of wino-like eigenstates, (ii) the non negligible branching ratios into the singlinos and singlet-like (pseudo)scalars. In particular, if the singlet states are much lighter/heavier than the SM Higgs boson, the dedicated electroweakino searches targeting the SM Higgs boson in the final state do not prove very efficient, since the signal regions are optimized for a signal where the scalar is compatible with a Higgs with mass $\approx125$~GeV.
	
	In our scan, the singlet (pseudo)scalars can have masses as low as a few GeV and up to 125 GeV (and beyond). We only focus on search channels for light (pseudo)scalars in the following. The phenomenology of light singlets in neutralino decays at the LHC have been discussed for many years \cite{Cheung:2008rh,Stal:2011cz,Cerdeno:2013qta,Kozaczuk:2015bea}. Light singlets can certainly appear in production channels like $\chi_1^\pm\chi_2^0$ with subsequent decays $\chi_1^\pm\rightarrow W^\pm \chi_1^0$ and $\chi_2^0\rightarrow \Phi\chi_1^0$ (with $\Phi$ a singlet-dominated scalar state). Viable final state configurations from associated chargino-neutralino production are $\left(\ell^\pm b\bar b,\, \ell^\pm\tau^+\tau^-,\,\ell^\pm\mu^+\mu^-,\,\ell^\pm\gamma\gamma \right)+\met$. We first discuss the signature with an isolated lepton, a photon pair and missing transverse momentum in the final state. The SM background rates are very small although this channel generally suffers from very low signal rates due to the small branching ratio. Nevertheless, ATLAS considered the diphoton channel in electroweakino pair production for a SM-like Higgs boson and demonstrated that this topology can be a viable LHC signature \cite{Aad:2015jqa}. In our scan, the diphoton branching ratio for a singlet-dominated state with mass below $125$~GeV can still reach the order of magnitude of that of the SM-like state. Thus, the branching-ratio suppression could still be balanced by the background-free aspect, for a singlet at, say, 80 GeV. Our light singlet scenarios could be probed in the diphoton and missing transverse momentum channel if signal regions with low invariant diphoton masses are introduced. For the mass-range between $20$ to $60$~GeV, the diphoton branching ratio is in general too low to be promising. Another regime is that of ultra-light singlets, with mass below two muon masses. Then, the loop-induced decay into photons can become dominant. However, the singlet also tends to become long-lived. Signatures with long-lived photons in GMSB-motivated scenarios have been studied in \cite{Aad:2014gfa}. Still, the mixing of a pseudoscalar with the neutral pion increases its lifetime, so that the decay of the pseudoscalar is prompt again in a small mass-window around $\approx m_\pi$ \cite{Domingo:2016unq,Domingo:2016yih}, but the photon pair will most likely appear as a single photon experimentally \cite{Domingo:2016unq}. GMSB-inspired searches targeting photon final states might then be sensitive \cite{Aaboud:2018doq}. However, we should stress that we did not obtain points in this extreme low-mass regime due to our scanning procedure. The reason is to be searched both in the limited number of points that we keep in the scan and the strong phenomenological constraints that apply on a very light (pseudo)scalar. In addition, the DM relic density calculation probably cannot be trusted in this mass-range, since it neglects hadronic effects, so that it made limited sense to look for such spectra actively.
	
	In the regime $2m_{\mu}<m_{\Phi}<2m_{\tau}$, the decay of the singlet into muons tends to dominate. Due to the small mass of the singlet, the muons would be very soft  $p_T\approx\mathcal{O}$(10) GeV and thus the threshold for the muon transverse momentum has to be set very low, at e.g.\ $7$~GeV. The leading lepton from the $W^\pm$ should be energetic enough to trigger the event. Moreover, a moderate cut on missing transverse momentum should help to further suppress the background. In contrast to the $\Phi\rightarrow\tau\tau$ channel, the invariant mass of the muon pair will reconstruct $m_\Phi$. However, for relatively heavy electroweakinos, the singlet will be highly boosted and thus both muons can be very collinear and might not be distinguishable. 
	
	Scenarios with $2m_\tau<m_\Phi<2M_B$ ($B$-meson mass) have chargino/neutralino topologies with a single light lepton, a tau pair and missing transverse momentum. This final state  might be very promising. SM backgrounds can be efficiently suppressed by demanding at least one light lepton and two hadronic taus, rejecting events compatible with a $Z$-boson. In addition, a cut on the transverse mass of the light lepton and missing transverse momentum further suppresses the SM background as in the generic MSSM multi-lepton electroweakino search focusing on SM Higgs with $H \rightarrow\tau\tau$ in final states. However, there is also a major difference due to the much lower singlet mass. The angular separation of both isolated taus originating from the same (pseudo)scalar, $\Delta R=\sqrt{\Delta\Phi^2+\Delta\eta^2}$, can be very small, i.e.\ both taus are almost collinear. In the worst case, separation might not even be possible. Moreover, the visible decay products of taus might be relatively soft since the neutrinos carry away a significant portion of the original tau energy. Despite all difficulties, Ref.~\cite{Cerdeno:2013qta} shows that signal isolation is possible. Ref.~\cite{Conte:2016zjp} investigated boosted ditaus signatures and estimated the sensitivity of boosted tagging techniques at the LHC.
	
	For $2M_B<m_\Phi\ll m_Z$, the $\ell b\bar b+\met$ final state is an interesting search channel. CMS \cite{Sirunyan:2017zss} considered this topology for a SM Higgs boson. They require the invariant mass of the two $b$-jets to be compatible with a parent SM Higgs boson and cut on the transverse mass as well as the contransverse mass. Signal regions are binned into various ranges of missing transverse momentum. A similar search strategy might work for our scenarios. Fig.~2 in Ref.~\cite{Sirunyan:2017zss} shows the SM $m_{b\bar b}$ distribution. It is clear that the main contribution comes from top pair production and the $m_{b \bar b}$ distribution peaks around 140 GeV. For much lower $m_{b\bar b}$ values, the background is steeply falling. Again, defining signal region with a large binning in a wide range of $m_{b\bar b}$ is mandatory.
	
	The $\chi_2^0\chi_3^0$ production channel with both eletroweakinos decaying into a singlet-like Higgs in association with missing transverse momentum can be a viable signal, e.g.\ $b\bar b b\bar b + \met$. Both ATLAS and CMS considered this signature in GMSB-motivated scenarios \cite{Aaboud:2018htj,Sirunyan:2017obz}, where the LSP is a massless gravitino and the pair produced neutral higgsinos decay purely into the SM Higgs bosons. Our benchmark points with light (pseudo)scalars cannot be probed by \cite{Aaboud:2018htj,Sirunyan:2017obz} due to the cuts on the invariant mass $m_{b\bar b}$ compatible with a SM Higgs boson. A large binning in a wide range of invariant masses of singlet candidates would improve sensititivy in such searches. 
	
	Actually the situation can become even more difficult as in so called stealth SUSY scenarios, which can be realised in the $Z_3$-violating NMSSM \cite{Ellwanger:2014hia}. A small mass-splitting between the NLSP and the scalar with $m_{\rm NLSP}\approx m_{\rm LSP}+m_{\Phi}$ and $m_{\rm LSP}\ll m_{\Phi},m_{\rm NLSP}$ can heavily reduce the missing transverse momentum, hence degrades the sensitivity at the LHC. Indeed, typical final states would involve hadrons (including hadronic tau decays) with little missing transverse momentum in which case the QCD backgrounds can be quite overwhelming. However, we stress that such configurations do not appear in our scans, first because, as explained in \cite{Ellwanger:2014hia}, such a scenario is difficult to realize in the $Z_3$-conserving case, and second because it is challenging to combine with the thermal relic density requirement (which typically demands a light scalar at twice the singlino mass).
	
	In the NMSSM, we have also encountered regions of the parameter space where, due to strong phase-space suppression, the decay $\chi_2^0 \rightarrow \chi_1^0 \gamma$ becomes dominant. This is particularly true in compressed configurations involving a singlino NLSP. Even for moderate mass splittings, the branching ratio can be large if the $Z$ decay-mode is suppressed due to the neutralino mixing matrices. ATLAS and CMS searches targeting final states with photons and missing transverse momentum might have sensitivity as it was shown in Ref.~\cite{Kim:2017pvm}. Such searches currently focus on GMSB-inspired scenarios, where the lightest electroweakino (typically a bino or wino) decays into a photon and the almost massless gravitino. However, the mass splitting is rather small in our relevant NMSSM scenarios $\Delta m=m_{\rm NLSP}-m_{\rm LSP}$ and for a large LSP mass, the photon only carries away little momentum. We explicitly tested a few benchmark points against dedicated GMSB search \cite{Aaboud:2018doq} but it seems that the mininum requirement on the transverse momentum of signal photons is rather severe and our benchmark points fail to pass those selection requirements, hence remain unconstrained. 
	
	Of course, it is likely that blind spots will persist in the electroweakino sector of the NMSSM but some additional coverage could be gained by considering more specific searches.
		
		\section{Conclusions}
		In this work, we considered electroweakino scenarios in the NMSSM that are characterized by light higgsinos and gauginos, with $|\mu|,\ |M_1|\text{ and/or } M_2\le500$~GeV, as well as possibly light singlinos and light singlet Higgs bosons. All squarks and sleptons decouple, hence avoiding direct search limits from ATLAS and CMS, and suppressing corresponding contributions to FCNC processes. In addition, the BSM doublet Higgs states are also chosen at a relatively high scale. We randomly generated NMSSM benchmark points and selected those satisfying the LEP, flavor and Higgs limits from \texttt{NMSSMTools}, as well as the upper bound on the thermal relic density of the LSP. However, we decided to discard constraints from direct DM searches, as these depend on additional assumption and our central aim is a collider analysis. Then, in view of testing the reach of the multi-lepton searches performed at the LHC, we considered five distinct NMSSM scenarios, namely, i) a MSSM-like scenario with no light singlinos or singlets, ii) the singlino-LSP scenario, iii) a scenario with singlino NLSP entering the decay chain of heavier electroweakinos, iv) a scenario including light Higgs singlet states mediating electroweakino decays and v) a scenario where a Higgs singlet has its mass on the annihilation threshold. For each benchmark point satisfying the limits mentioned above, we generated MC event samples and handed them to {\tt CheckMATE}, which tests a benchmark scenario against current ATLAS and CMS SUSY searches. We considered a selection of relevant electroweakino searches, covering a large class of electroweakino final state topologies. All these searches have been implemented and fully validated in {\tt CheckMATE}. As expected, the multilepton ($\ge3\ell$)-searches show the best sensitivity while the soft dilepton searches lead to a partial coverage in the coannihilation region.
		
		The main results are summarized in Fig.~[1-5]. In these, a model point is clearly excluded only if the predicted number of signal events is larger than $1.5$ times the $95\%$ C.L. upper bound, while clearly allowed points have a predicted signal that is at least a factor $0.67$ below the nominal bound for the most sensitive signal region. Large regions of parameter space of the electroweakino NMSSM scenarios which were allowed by 8 TeV data are now covered by Run-2 data. However, many benchmark scenarios with very light neutralinos are still allowed by the 2016-2017 data of Run-2 of the LHC, which should conclude data collection at the end of 2018.
		
		The standard search channels for weakly-interacting particles at the LHC primarily rely on energetic light leptons in the final states. We have seen that these channels continue to be relevant for the chargino-neutralino sector of the NMSSM. However, we also note that the comparatively small alteration with respect to the electroweakino sector of the MSSM is sufficient to highlight some new effects. From the perspective of the relic density, the LSP annihilation in the singlet Higgs funnels opens a large pannel of kinematical configuration which could affect the collider searches. In addition, the existence of a singlino state implies a possible new ladder in the decays of SUSY particles, which could also modify the multiplicity of the leptons in the final state and their energy. Finally, the possibility of decays involving light Higgs states tends to strengthen the final states with $\tau$'s, which are more difficult to identify. Admittedly, these features appear more as curiosities than leading trends in the NMSSM electroweakino phenomenology, but they can open `exceptions' in the exclusion picture of collider searches. In fact, we observe that the pattern of constraints, even in configurations with light spectra, is far from homogeneous and that, at least from the statistical approach of \texttt{Checkmate}, many points remain allowed in kinematical configurations that are naively excluded in the `idealized' scenarios. 
		
		Obviously, increased statistics should achieve the exclusion of many points that are already constrained. However, considering that the mechanisms that allow some spectra to evade limits from light-lepton searches are structural, it is likely that many of them will continue to be resilient to this form of searches. Searches considering $b\bar{b}$ or $\tau\tau$ pairs from a light Higgs state could help improve the coverage of these scenarios. In addition, compressed configurations with sizable $\chi_2^0\to\chi_1^0\gamma$ are easily achieved in the NMSSM, since both the bino and the singlino can be comparatively light and that the singlino naturally mixes with the higgsinos. Photonic searches, similar to those currently considered in GMSB-inspired frameworks but targeting soft photons, can be expected to cover this type of phenomenologies.	
		
		\section*{Acknowledgements}
		We acknowledge discussions with David G.\ Cerde\~no in the initial stages of this project. We thank Krzysztof Rolbiecki for useful comments.
		
        F.\ D.\ and  P.\ M.-R.\ acknowledge support from the Spanish Research Agency (``Agencia Estatal de Investigaci\'on'') through the contract FPA2016-78022-P and IFT Centro de Excelencia Severo Ochoa under grant SEV-2016-0597. In addition, the work of F.\ D. was supported in part by the DFG Research Group ``New Physics at the LHC'' Project D. V.\ M.\ L. acknowledges support of the BMBF under project 05H18PDCA1.  The work of R.\ RdA. has been supported by MINECO, Spain, under contract FPA2014-57816-P and Centro de excelencia Severo Ochoa Program under grants SEV-2014-0398, by the European Union projects H2020-MSCA-RISE-2015-690575-InvisiblesPlus and H2020-MSCA-ITN-2015/674896-ELUSIVES and by Generalitat Valenciana grant PROMETEOII\-2014/050.

	\end{document}